%% file: Fpaper.tex
\documentclass[11pt]{article}

\usepackage{verbatim}
\usepackage{amsmath}
\usepackage{amsfonts}
\usepackage{latexsym}
\usepackage{epsfig}
\usepackage{color}
\if@twoside\oddsidemargin25mm
\else
\input{texdefs}

\begin{document}

\input{Ftitle}

\setcounter{page}{1}

\input{Fintro}

\input{Ffieldtadp}

\input{Fstringtadp}

\input{Fshape}

\input{Fconcl}

\appendix 
\def\theequation{\thesection.\arabic{equation}}

\bibliographystyle{paper}
{\small
\bibliography{paper}
}

\end{document}

%% file: texdefs.tex
\renewcommand{\d}{\mathrm{d}}

\DeclareMathSymbol{\mg}{\mathrel}{symbols}{"1D}

\newcommand{\gd}{\delta}

\newcommand{\gf}{\phi}

\newcommand{\gr}{\rho}

\newcommand{\gt}{\tau}

\newcommand{\gp}{\pi}

\newcommand{\gD}{\Delta}

\newcommand{\gL}{\Lambda}

\newcommand{\gTh}{\Theta}

\newcommand{\cE}{{\cal E}}
\newcommand{\cF}{{\cal F}}
\newcommand{\cG}{{\cal G}}

\newcommand{\cT}{{\cal T}}

\newcommand{\ui}{{\underline i}}
\newcommand{\uj}{{\underline j}}

\newcommand{\tr}{\text{tr}}

\newcommand{\ra}{\rightarrow}

\newcommand{\inv}{^{-1}}
\newcommand{\ovr}[1]{{\overline{#1}}}

\newcommand{\labl}[1]{\label{#1}}
\newcommand{\beq}{\begin{equation}}
\newcommand{\eeq}{\end{equation}}
\newcommand{\barr}{\begin{array}}
\newcommand{\earr}{\end{array}}
\newcommand{\equ}[1]{\begin{gather} #1 \end{gather}}

\newcommand{\tabu}[2]{\begin{tabular}{#1} #2 \end{tabular}}
\newcommand{\arry}[2]{\begin{array}{#1} #2 \end{array}}

\newcounter{oldcounter}

\newcommand{\bgt}{{\bar\tau}}

\newcommand{\Intr}{\mathbb{Z}}
\newcommand{\Cplx}{\mathbb{C}}

\newcommand{\ba}[2]{\[\begin{array}{#2}\label{#1}}
\newcommand{\ea}{\end{array}\]}
\newcommand{\be}{\begin{equation}}
\newcommand{\ee}{\end{equation}}
\newcommand{\bea}{\begin{eqnarray}}
\newcommand{\eea}{\end{eqnarray}}

\newcommand{\E}[1]{\mathrm{E_{#1}}}
\newcommand{\U}[1]{\mathrm{U(#1)}}
\newcommand{\SU}[1]{\mathrm{SU(#1)}}
\newcommand{\SO}[1]{\mathrm{SO(#1)}}

\newcommand{\rep}[1]{\mathbf{#1}}
\newcommand{\crep}[1]{\overline{\rep{#1}}}

%% file: Ftitle.tex
\thispagestyle{empty}

\begin{flushright}
FTPI-MINN-03/30 \\
UMN-TH-2219/03  \\      
hep-th/0311015
\end{flushright}
\vskip 2 cm
\begin{center}
{\Large {\bf 
Stringy profiles of gauge field tadpoles near orbifold singularities: 
\\[2ex]
II.\ field theoretical implications 
}
}
\\[0pt]
\vspace{1.23cm}
{\large
{\bf Stefan Groot Nibbelink$^{a,}$\footnote{
{{ {\ {\ {\ E-mail: nibbelin@hep.umn.edu}}}}}}}, 
{\bf Mark Laidlaw$^{b,}$\footnote{
{{ {\ {\ {\ E-mail: mlaidlaw@perimeterinstitute.ca}}}}}}},
\bigskip }\\[0pt]
\vspace{0.23cm}
${}^a$ {\it 
William I. Fine Theoretical Physics Institute, 
School of Physics \& Astronomy, \\
University of Minnesota, 116 Church Street S.E., 
Minneapolis, MN 55455, USA. \\}
\vspace{0.23cm}
${}^b$ {\it  
Perimeter Institute for Theoretical Physics, \\
35 King Street North, Waterloo, Ontario N2J 2W9, Canada
}
\bigskip
\vspace{1.4cm} 
\end{center}
\subsection*{\centering Abstract}

Recent work has provided a direct string calculation of the internal
coordinate dependence of gauge field tadpoles on the orbifold
$\Cplx^3/\Intr_3$. We investigate the structure of these profiles in
momentum and 
coordinate space representations using both analytical and numerical
methods. The twisted states are determined to be localized within a few
string lengths of the singularity. This provides an example of how
delta--like singularities, which are typical for field theories on
orbifolds, are avoided in string theory. A systematic expansion
of the full string result allows us to speculate on how 
the leading stringy effects can be incorporated in a field theoretical
description. In one heterotic model, we find that even
though tachyons are not present in the physical spectrum, they 
totally dominate the tadpole profile in the vicinity of the fixed
point.

\newpage

%% file: Fintro.tex
\section{Introduction}
\labl{sc:intro}

In this second paper in a series of two on the profiles of gauge field 
tadpoles in extra dimensions string and field theory, we offer a field
theoretical interpretation of the string results obtained in ref.\
\cite{GNL_I}.  As opposed to that paper, which was written for the
string oriented audience, the presentation in this paper is aimed at 
those interested  in physics of extra dimensions and string
phenomenology.

In the recent literature there has been a great deal of attention to the
study of theories with  extra dimensions,  some inspired by the works of 
refs.\ \cite{Randall:1999ee,Randall:1999vf,Arkani-Hamed:1998rs,
Barbieri:2000vh}. Many of the proposed models are based on 
simple orbifolds like $S^1/\Intr_2$ or $S^1/\Intr_2\times \Intr_2'$. 
It is not entirely clear that field theories on orbifolds are
physically well--defined: Near the orbifold singularity,  strong
curvature effects may lead to a breakdown of field theory
computations. Moreover, theories of extra dimensions are in general
nonrenormalizable, and should therefore be defined with some sort of
cut--off. In this context, calculations that involve momenta much
larger than this cut--off are beyond the validity of the theory. On  
the other hand, to probe the full structure of orbifold singularities
arbitrary momenta are needed. A simple momentum cut--off also does not
seem to be appropriate, since it breaks Lorentz invariance and may
violate other defining symmetries of the orbifold field theory. The
only consistent ultra--violet completion of field theory we have to
date is string theory. It evades  this problem by providing the string
scale that is the natural cut--off scale for field theory
\cite{Polchinski:1986zf,Ghilencea:2001bv}, while maintaining important
symmetries like Lorentz invariance at the same time.

Fayet--Iliopoulos tadpoles on orbifolds provide a particularly
interesting setting in which such consistency questions can be put
forward. Four dimensional Fayet--Iliopoulos $D$--terms
\cite{Fayet:1974jb} have a natural generalization to five dimensional
supersymmetric gauge theories on orbifolds. As observed in 
ref.\ \cite{Mirabelli:1998aj} the effective auxiliary field at the
boundaries of $S^1/\Intr_2$ also contains a derivative with respect to
the fifth dimension of a real scalar field part of the five dimensional
gauge multiplet. There have been several calculations showing that
such interactions are generated at the one--loop level
\cite{Ghilencea:2001bw,Barbieri:2001cz,Scrucca:2001eb,
GrootNibbelink:2002wv,GrootNibbelink:2002qp}. For bulk states the result
is given by a massive quadratically divergent integral times
delta--functions at the orbifold singularities. Here the ``mass''
represents the derivative squared  with respect to the fifth dimension
acting on  these delta--functions. By investigating the required
counter--term structure for this real scalar field of the gauge
multiplet, it has been argued in refs.\ 
\cite{GrootNibbelink:2002wv,GrootNibbelink:2002qp,Lee:2003mc} 
that these tadpoles may lead to strong localization of bulk zero
modes. This treatment has not been universally accepted, for
alternative treatments see refs.\  \cite{Barbieri:2002ic,Marti:2002ar}.

The existence of Fayet--Iliopoulos tadpoles for both auxiliary and
derivative of physical fields is in no way particular to five
dimensional models. One--loop computations in 
higher dimensional field theoretical models 
\cite{vonGersdorff:2002us,Csaki:2002ur} and in field theory
approximation of heterotic string models  \cite{GrootNibbelink:2003gb}
have confirmed that these tadpoles are generated locally at orbifold
fixed points.  (Fayet--Iliopoulos $D$--terms for zero modes have been
discussed in heterotic \cite{Dine:1987xk,Atick:1987gy,Dine:1987gj} and
type I string \cite{Poppitz:1998dj} context.) 
These  results provided the main motivations to 
establish that such tadpoles also arise in string theory \cite{GNL_I}. 
The main objective of this paper is to compare the previously obtained
field theory results for gauge field tadpoles with our string calculations. 
To this end we perform both analytical and numerical analyses
of the string expressions to compare them to the field theoretical
expectations. We seek to give a physical  interpretation of the
stringy corrections.

This paper is organized as follows: We begin in section
\ref{sc:fieldtadp} by reviewing the low energy field theory description
of gauge field tadpoles in heterotic models. In section
\ref{sc:stringtadp} we describe our results for these tadpoles obtained
by direct string calculation \cite{GNL_I} in a language chosen to make 
the similarities and differences with the field theory results manifest. 
Section \ref{sc:shapes} is devoted to detailed studies of the resulting
profiles of the gauge field tadpoles. We investigate which states give
leading contributions: Surprisingly,  we find that tachyons
sometimes play an important role. Finally, we discuss some
implications of our results in section \ref{sc:concl}.

%% file: Ffieldtadp.tex
\section{Gauge field tadpoles in field theory}
\labl{sc:fieldtadp}

In this section we recall some results concerning localized tadpoles
in the heterotic theory obtained using field theory methods. It is
based to a large extent on the results obtained in  
refs.\ \cite{GrootNibbelink:2003gb,Gmeiner:2002es}. 
As we are primarily interested in local properties, we have chosen to 
work on the non--compact orbifold $\Cplx^3/\Intr_3$ instead of the
compact orbifold $T^6/\Intr_3$: This avoids the introduction of
another scale that obscures the tracing of the genuine stringy effects. 
The orbifold action $\gTh$ leads to diagonal identifications 
\equ{
\gTh ~:~ z_i \ra  e^{2\gp i\, \gf_i} \, z_i  
}
of the complex coordinates  $z = (z_i) = (z_1,z_2,z_3)$. 
We consider the low--energy spectrum of the heterotic 
$\E{8} \times \E{8}'$ string on this orbifold, consisting of
a ten dimensional $N=1$ supergravity multiplet and a super
Yang--Mills theory with the corresponding gauge group. At the orbifold
fixed point the spectrum of the theory is projected to a four dimensional 
$N=1$ supersymmetric theory. When we take non--trivial boundary
conditions for the action of the orbifold twist on the ten dimensional
gauge field one--form 
\equ{
A(x, z) \ra A(x, \gTh\, z) = U\, A(x, z)\, U\inv,
\qquad 
U = e^{2\gp i\, v_a^I H_{a}^I}, 
}
we obtain a four dimensional gauge theory with a reduced gauge group
$G$ and some specific four dimensional chiral matter representations
$\rep{R}$. Here the generators $H^I_1$ and $H^I_2$ denote the 
$\SO{16} \subset \E{8}$  and $\SO{16}' \subset \E{8}'$ Cartan
elements, respectively. The chiral matter $\rep{R}$ always comes in
three identical copies, because there are three internal gauge fields
$A_\ui^{\rep{R}}$ in the representation of the gauge group. In addition,
there are four dimensional twisted states at the orbifold fixed point
that also form  chiral multiplets. These twisted states appear in two
varieties: some come in three equal copies (their representation is
denoted by $\rep{T}$), while others are singlets (referred to as
$\rep{S}$). The full spectrum of the theory at the fixed point can be
shown to be equivalent to one of those given in table
\ref{tab:z3models}.

\begin{table}
{\small 
  \begin{center}
\renewcommand{\arraystretch}{1.25}
  \begin{tabular}{|l|l|l|ll|l}\hline
    {Model }
      & {Shift $(v_1^I ~|~ v_2^I)$ and }
      & {Untwisted}
      & {Twisted}
      & 
\\
& {gauge group $G $} 
& {$(\rep{3}_H,\rep{R})$ }
& {$(\rep{1}_H, \rep{S})$ } 
& {$(\rep{\bar{3}}_H, \rep{T})$}
      \\\hline
    $\E{8}$
      & $\frac{1}{3}\!\left(~0^8 ~~~~ ~~~~~ ~~|~~ 0^8 ~~~~~ ~~~~~ \right)$
      & 
      &  
      & $(\rep{1})(\rep{1})'$
      \\
      &    $~~~~~~ ~~~~~~ ~\, \E{8}  \times   \E{8}'$
      & & & 
      \\\hline
    $\E{6}$
      & $\frac{1}{3}\!\left(\mbox{-} 2,~1^2,~ 0^5 ~~|~~ 0^8 ~~~~~ ~~~~~ \right)$
      & $(\rep{27},\crep{3})(\rep{1})'$
      & $(\rep{27},\rep{1})(\rep{1})'$
      & $ (\rep{1},\rep{3})(1)'$
      \\
       & $~~~ \E{6}\!\times\!\SU{3} \times \E{8}' ~~~~~~~ $
      & & &
      \\\hline
    $\E{6}^2$
      & $\frac{1}{3}\!\left(\mbox{-}2,~1^2,~0^5 ~~|~~ \mbox{-}2,~1^2,~0^5\right)$
      & $(\rep{27},\crep{3})(\rep{1},\rep{1}) \!+\! (\rep{1},\rep{1})(\rep{27},\crep{3})'$
      & $(\rep{1},\rep{3})(\rep{1},\rep{3})'$
      &  
      \\
      & $~~~ \E{6}\!\times\!\SU{3} \times \E{6}'\!\times\!\SU{3}'$
      &  & & 
      \\\hline
    $\E{7}$
      & $\frac{1}{3}\!\left(~0,~1^2,~0^5 ~~|~~ \mbox{-}2,~0^7 ~~~~~ \right)$
      & $(\rep{1})_{0}(\rep{64})'_{\frac{1}{2}}
 + (\rep{56})_{1}(\rep{1})'_{0}$
      & $(\rep{1})_{\frac{2}{3}}(\rep{14})'_{\mbox{-}\frac{1}{3}} $
      & $ (\rep{1})_{\frac{2}{3}}(\rep{1})'_{\frac{2}{3}}$
      \\
      &  $ ~~~~\,  \E{7}\!\times\!\U{1} \times
\SO{14}'\! \times\! \U{1}' \!\!$
      & $+\, (\rep{1})_{0}(\rep{14})'_{\mbox{-}1} + (\rep{1})_{\mbox{-}2}(\rep{1})'_{0}$
      & $ +\,  (\rep{1})_{\mbox{-}\frac{4}{3}}(\rep{1})'_{\frac{2}{3}}$ 
      & 
      \\\hline
    $\SU{9}$
     & $ \frac{1}{3}\! \left(\mbox{-}2,~1^4~,0^3 ~~|~~ \mbox{-}2,~0^7
~~~~~ \right) $
     & $(\rep{84})(\rep{1})'_{0} + (\rep{1})(\rep{64})'_{\frac{1}{2}}$
      & $(\crep{9})(\rep{1})'_{\frac{2}{3}}$
     & 
      \\
      & $~~~~~ ~~~~ \SU{9} \times \SO{14}'\!\times\!\U{1}' \!\!$
      &$+\, (\rep{1})(\rep{14})'_{\mbox{-}1}$ & &
      \\\hline
    \end{tabular}
  \end{center}
}
  \caption{
The spectra at the fixed point of the five $\Intr_3$ orbifold models
are displayed.   
 }  
  \label{tab:z3models}
\end{table}

Of the five possible models listed in table \ref{tab:z3models} only
the last two have $\U{1}$'s. The generators $q_a = v_a^I H_{a}^{I}$
are potentially anomalous: It can be shown that $\U{1}$ generated by
$q_1$ is anomalous in the $\E{7}$ model, while it is not in the
$\SU{9}$ model. However, in that model the charge $q_2$ is
anomalous. These $\U{1}$'s are only anomalous at the orbifold fixed
point \cite{Gmeiner:2002es}, and as  was investigated in 
\cite{GrootNibbelink:2003gb} they can be canceled through a local
version of the Green--Scharz mechanism \cite{Green:1984sg}. An
anomalous $\U{1}$ is expected to be spontaneously broken by a
Fayet--Iliopolous tadpole for auxiliary $D$--field of the anomalous
vector multiplet \cite{Dine:1987xk}. In ref.\
\cite{GrootNibbelink:2003gb} the shape of this tadpole on the orbifold
$T^6/\Intr_3$ was calculated, and it was shown that similar tadpoles
are generated for the internal gauge fields $A_j^a$ corresponding to
these $\U{1}$'s. Using the methods discussed in
\cite{GrootNibbelink:2003gd} for doing field theoretical calculations
on orbifolds these results are readily extended to the non--compact
orbifold $\Cplx^3/\Intr_3$ as well. In ten dimensional momentum space
the result reads  
\equ{
\langle A_j^a (k_6) \rangle 
= 
i \, k_{\uj} \frac{\gd^4(k_4) }{(2\pi)^4} \int \frac {d^4 p_4}{(2\gp)^4} 
\left\{ 
\frac {3\, \tr_{\rep{R}}(q_a)}{27} 
\frac 1{p_4^2 + \frac 13 k_i k_{\ui}} 
+ 
\Bigl( 3 \,\tr_{\rep{T}}(q_a) + \tr_{\rep{S}}(q_a) \Bigr) 
\frac 1{p_4^2}
\right\}  .
\labl{fieldtheorytadpole} 
}
The trace of charge $q_a$ over representation $\rep{r}$ is denoted by
$\tr_{\rep{r}}(q_a)$. The four dimensional delta $\gd^4(k_4)$ is a
consequence of four dimensional momentum conservation. 
All complex internal momenta $k_6 = (k_i, k_\ui)$ are allowed, 
because this expression is the Fourier transform of (derivatives
of) the delta function $\gd^6(z)$ at the orbifold fixed point. 
The four dimensional momentum integrals are clearly divergent, and
in particular  the leading (quadratic) divergence is proportional to
the trace of the charges over the full spectrum at the fixed point.
This quadratic divergence is present only when there is an anomalous
$U(1)$ in the spectrum \cite{Dine:1987xk}.

The four dimensional divergences can be regularized in various ways.
We follow \cite{Polchinski:1986zf}  and use a variant of Schwinger's
proper time regularization in order to make contact with the string 
calculation of the next section.  We write 
\equ{
\int \frac{\d^4 p_4}{p_4^2 + m^2} = 
 \frac {\gp}4 \gL^2\, \int_{-\frac 12}^{\frac 12} \d \gt_1 \, 
\int_{1}^\infty \frac{\d \gt_2}{\gt_2^2}\,  
e^{-4\gp\, \gt_2\, m^2/\gL^2},
\labl{Schwinger}
}
for an arbitrary mass $m$ and a cut--off scale $\gL$. The $\gt_1$
integration is clearly redundant, but it has been included in this formula
for the subsequent comparison with the string theory tadpole. Using
the proper time representation of the momentum integrals the
expression  for the tadpole may be written as:
\equ{
\langle F_{j\uj}^b (k) \rangle 
= 
\frac{\gd^4(k_4)}{(2\gp)^4} \, 
\frac {\gp}4 \gL^2 \, 
\int\limits_{\mbox{-}\frac 12}^{\frac 12} \d \gt_1
\int\limits_{1}^\infty \frac {\d \gt_2}{\gt_2^2} \, 
\sum_{s = un, tw}\, Q_s^b \, 
e^{- \gD_s\, k_i k_\ui /\gL^2}
.
\labl{TadFieldSchwinger}
}
Here we have combined tadpoles for $\smash{A_j^b}$ and
$\smash{A_\uj^b}$ after a partial integration to obtain the field
strength $\smash{F_{j\uj}^b}$. The quantities 
$\smash{Q_{un}^b, Q_{tw}^b, \gD_{un}}$ and
$\smash{\gD_{tw}}$ appearing in this expression are given by 
\equ{
Q_{un}^b = \frac {3}{27}\,  \tr_{\rep{R}}(q_b),
\quad 
Q_{tw}^b =  \tr_{\rep{S}}(q_b) + 3 \,\tr_{\rep{T}}(q_b), 
\qquad  
\gD_{un} = 4\gp\, \gt_2\, \frac 13, 
\quad  
\gD_{tw} =0. 
\labl{QgDfield}
}
The subscripts $un$ and $tw$ refer respectively to the untwisted and 
twisted sectors. The width $\gD_s$ could in principle carry internal
space indices $i$ or $\ui$, but rotational symmetry of the orbifold
requires them all to be the same. For the untwisted sector 
$\gD_{un}\neq 0$ and the tadpole takes the form of Gaussian
distributions with widths depending on $\gt_2$. This variable is 
integrated up to the cut--off of the effective field theory, which has
be scaled to unity by pulling out $\gL$ explicitly in equation
\ref{Schwinger}. In particular for large internal momenta $k_6$ the tadpole
will be damped considerably. For the fixed point states there is no
such suppression because $\gD_{tw} = 0$. This reflects the fact
that in field theory we have treated the twisted states as fields
that are localized exactly at the fixed points. In the next section we
will see that the string theory treatment of these states is
fundamentally different.

In order to have an exact measure of when and to what extent
states are localized we introduce the concept of  localization width, 
defined as 
\equ{
\ovr{\gD_s} = 
\int\limits_{\mbox{-}\frac 12}^{\frac 12} \d \gt_1
\int\limits_{1}^\infty \frac {\d \gt_2}{\gt_2^2} \, 
 {\gD_s(\gt_1,\gt_2)}.
\labl{LocWidthF}
}
One can think of this quantity as the average radial spread of a given
state around the orbifold fixed point. 
A bulk, or delocalized, state has a (logarithmically) divergent
localization width, while this width is zero for a fixed point
state. It is not difficult to see that this definition precisely
matches with the interpretation of the untwisted and twisted states in
the field theory treatment, since we obviously have 
$\ovr{\gD_{un}} = \infty$ and $\ovr{\gD_{tw}} = 0$, respectively.

%% file: Fstringtadp.tex
\section{Gauge field tadpoles in string theory}
\labl{sc:stringtadp}

The gauge field tadpoles, which were described from a field theory
point of view, also arise at one loop in the full heterotic string
theory. In this paper we wish to compare the field and string theory
results to each other, therefore we discuss the schematic structure
of the string theory tadpoles here. Details of the string computation
are given in ref.\ \cite{GNL_I}. Here we only call to the attention of
the reader those ingredients  that are essential to the
understanding of the analysis of the tadpoles that we present below.

A one loop calculation of a tadpole in closed string theory involves
averaging the corresponding vertex operator over inequivalent torus
world sheets. A  two dimensional torus is defined as the complex plane
$\mathbb{C}$ divided by the lattice spanned by $1$ and the 
complex number $\gt = \gt_1 + i \gt_2$. Only by restricting this
(modular) parameter $\gt$ to a fundamental domain, the inequivalent
tori are labeled uniquely. In figure \ref{fg:ModularPara} we have
depicted the fundamental domain $\cF$ which is defined by $|\gt| \geq 1$ 
and $-\frac{1}{2} \le \gt_1 \le \frac{1}{2}$.  This corresponds to the
most common choice for the fundamental domain in string theory.
The field theory cut--off at the string scale would be given by 
$-\frac{1}{2} \le \gt_1 \le \frac{1}{2}$ and $\gt_2 \geq 1$, as we
have discussed in the previous section.

\begin{figure}
\[
\arry{ccc}{
\raisebox{2ex}{\scalebox{0.9}{\mbox{\input{torus.pstex_t}}}}
& \qquad \qquad & 
\raisebox{0ex}{\scalebox{0.8}{\mbox{\input{FunDom.pstex_t}}}}
\\[2ex] 
\text{world sheet torus} && 
\cF = \Bigl\lbrace 
\gt \in \Cplx \, | \, \mbox{-}\frac 12 < \gt_1 < \frac 12, ~
\gt_2 > \sqrt{ 1 - \gt_1^2 }
\Bigr\rbrace, 
}
\]
\caption{
The first picture gives a torus defined by the complex variable 
$\gt$. To only label inequivalent tori this parameter is restricted to
the fundamental domain $\cF$, depicted in the second picture. 
\labl{fg:ModularPara}
}
\end{figure}
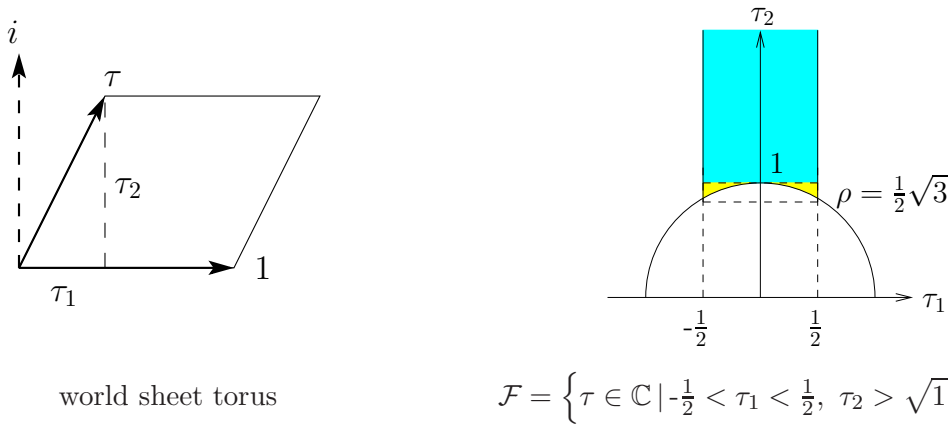

The $\Intr_3$ orbifolding leads to a classification of string states
in several sectors which are determined by the periodicities of the
coordinate fields under the lattice shifts $1$ and $\gt$.  These can
be characterized broadly as untwisted and twisted sectors. The
untwisted states (the $u$--sector) have trivial boundary conditions
under the shift $1$.  We introduce a finer classification of the
twisted states that distinguishes between those with trivial boundary 
conditions under the shift $\gt$ (the $t$--sector) and those with 
non--trivial boundary conditions in both directions (the
$d_\pm$--sectors).  The distinction between these doubly twisted
states is whether they have the same or opposite orbifold conditions
under both lattice shifts. In our classification the $u$--sector
corresponds to the untwisted field theory states ($un$), and the $t$--
and $d_\pm$--sectors are the twisted states ($tw$). The gauge field
tadpole takes the form of a sum over these different sectors 
\equ{
\langle F_{j\uj}^{b}(k_6) \rangle = 
\frac{\gd^4(k_4)}{(2\gp)^4}\,  \sum_{s = u, t, d_\pm} G^b_s(k_6), 
\quad 
G^b_s(k_6) = 
\int_\cF  \frac {\d^2 \gt}{\gt_2^2}
\, Q_s^{b}(\gt) \, 
e^{- \gD_s(\gt,\bgt) \, k_i k_\ui }. 
\labl{SketchTadp}
}
In this expression $Q_s^b(\gt)$ can be thought of as the trace of the
$q_b$ charges of sector $s$. In string theory the orbifold space
coordinates $X^i$ and $X^\ui$ are fields on the torus world
sheet. Their expectation value 
\(
\gD_s(\gt,\bgt) = \langle X^\ui X^i\rangle_s(\gt, \bgt) 
\)
set the widths of the orbifold singularity for the states of sector
$s$. This interpretation is motivated by the observation that these 
expectation values appear as the (inverse) standard deviations of the
Gaussian distributions for the internal momentum $k$ of the gauge
field.  (Here we have dropped the subscript on the internal momenta 
$k_6$.) Rotational symmetry requires that the widths $\gD_s(\gt,\bgt)$ 
are equal in all three complex momentum directions for each sector. 
The exact one loop string expression for the functions $Q^b_s(\gt)$
and $\gD_s(\gt,\bgt)$ have been obtained in ref.\ \cite{GNL_I}.  The
notation of this expression \eqref{SketchTadp} emphasizes that
$Q^b_s(\gt)$ is holomorphic in $\gt$ and that $\gD_s(\gt, \bgt)$ is a
real function of both $\gt$ and $\bgt$. For this reason it is
necessary to take care when interpreting these quantities (in field
theory).  We emphasise that even though $Q_s^b(\gt)$ is a complex
function, the tadpole \eqref{SketchTadp} is real, because the
integration over the fundamental domain is symmetric  
under $\gt_1 \ra -\gt_1$.

The field theoretical expression \eqref{TadFieldSchwinger} for the gauge
field tadpole has a structure compatible with that of string theory
\eqref{SketchTadp}. But even at this crude level some important
differences can be identified: In field theory we only distinguish between
untwisted ($un$) and twisted states ($tw$); the string classification
is finer since the twisted states are now divided up into single ($t$)
and doubly twisted ($d_\pm$) states, which have different profiles over 
the extra dimensions as we will demonstrate in detail in section
\ref{sc:shapes}.  In addition, the integration over $\gt_1$ and $\gt_2$ is
different: In string theory the lower bound of $\gt_2$ depends on
$\gt_1$, because of the shape of the fundamental domain given in
figure \ref{fg:ModularPara}, while in field theory it is equal to
unity. Neglecting the peculiar shape of the fundamental domain
gives an error of the order of  $\gp/3 -1 \approx 0.047$ 
(taking the factor $1/\gt_2^2$ as part of the integration
measure). There is no explicit cut--off dependence in the string
result because we have chosen string units: $\gL = M_s = 1$.

The tadpole takes the form in momentum space of a sum of 
Gaussian distributions with $\gt$ dependent standard deviations
$\gD_s$, it is straightforward to obtain the Fourier transform to
the internal coordinate space 
\equ{
\langle F_{j \uj}^{b}(z) \rangle = 
\frac{\gd^4(k_4)}{(2\gp)^4}\,   
\sum_{s=u,t,d_\pm} \cG^b_s(z), 
\quad 
\cG^b_s(z) = 
\int_\cF  \frac {\d^2 \gt}{\gt_2^2}\, 
Q_s^{b}(\gt) \, 
\Bigl( \frac{1 }{ \gp\gD_s(\gt, \bgt)} \Bigr)^3
\,e^{-  \bar z z / \gD_s(\gt,\bgt)}. 
\labl{SketchTadpCoord}
}
It should, however, be stressed that the Fourier transform only exists
when all $\gD_s(\gt) \geq 0$ for all $\gt$.

To gain more insight into  the relationship  between the field and string
predictions for the gauge field tadpoles, we systematically
approximate the string results to an accuracy that is roughly of the 
same order as the field theory approximation of the fundamental domain
of the string. The functions $Q_s(\gt)$ and $\gD_s(\gt,\bgt)$ can both
be written as a power series in $\exp(-2 \pi \gt_2)$. Due to the
$\Intr_3$ orbifolding  fractional powers $1/3$ and $2/3$ of
this exponential factor also arise. Since $\gt$ is restricted to
take values within the fundamental domain, 
$\gt_2 \geq \gr = \frac 12\sqrt{3}$, and 
consequently the exponential factor $\exp(-2 \pi \gt_2)$ always gives a
sizable suppression even when $\gt_2$ is not very large. We
have evaluated those exponentials for various small values of $\gt_2$
in table \ref{tb:ValuesExp}: The table shows that corrections of the
order of $\exp(-2 \pi \gt_2)$ are at least suppressed by a factor
$0.004$, which is an order of magnitude smaller than the factor of 
$0.05$ associated with the field theory motivated truncation of the 
fundamental domain.

\begin{table}
\[
\renewcommand{\arraystretch}{1.5}
\arry{| l | l l l l |}{
\hline 
\multicolumn{1}{|c}{\gt_2 } & \multicolumn{1}{|c}{\frac 12 \sqrt 3} 
& \multicolumn{1}{c}{1} & \multicolumn{1}{c}{1.5} 
&    \multicolumn{1}{c|}{2} 
\\ \hline 
e^{- 2\pi \, \frac 13 \, \gt_2}  & 
0.163 & 0.123 & 0.043 & 0.015 
\\
e^{- 2\pi \, \frac 23 \, \gt_2}  & 
0.027 & 0.015 & 0.002 & 0.000\, 2 
\\ 
e^{- 2\pi \,\, \gt_2}  & 
0.004 & 0.002 & 0.000\, 1 & 0.000\, 03 
\\ \hline 
}
\]
\caption{
In this table we have tabulated the values of first few exponential
factors in $\gt_2$ that arise in the expansions of $Q_s(\gt)$ and
$\gD_s(\gt,\bgt)$ for a $\Intr_3$ orbifold. As these factors are quite
small, it is to be expected that the first few terms give a reasonably
accurate approximation of the full string results. 
\labl{tb:ValuesExp} 
}
\end{table}

Motivated by this, we expand the exact expressions for
$\gD_s(\gt,\bgt)$ obtained in ref.\ \cite{GNL_I} to order 
$\exp(-2 \pi\gt_2)$:  
\equ{
\arry{lcl}{
\gD_u(\gt,\bgt) &= &
c - 3 \ln 3 + 4 \gp  \gt_2 \frac {1}{3} + \ldots, 
\\[2ex]
\gD_t(\gt,\bgt) &= &
c 
+ 6 \cos(2 \gp\frac {\gt_1}3) e^{- 2 \pi \gt_2 \, \frac 13}
+ 9 \cos(4\gp \frac {\gt_1}3) e^{-2\pi \gt_2\,\frac 23}+ \ldots,
\\[2ex]
\gD_{d_+}(\gt,\bgt) &= &
c  
+ 6 \cos(2 \gp \frac {\gt_1-1}{3}) e^{- 2 \pi \gt_2\, \frac 13}
+ 9 \cos(4 \gp \frac {\gt_1-1}{3}) e^{- 2 \pi \gt_2\, \frac 23}
+ \ldots, 
\\[2ex]
\gD_{d_-}(\gt,\bgt) &=  &
c  
+ 6 \cos(2 \gp \frac {\gt_1+1}{3}) e^{- 2 \pi \gt_2\, \frac 13}
+ 9 \cos(4 \gp \frac {\gt_1+1}{3}) e^{- 2 \pi \gt_2\, \frac 23}
+ \ldots.
}
\labl{ApprxgD}
}
The constant $c$ is related to the normal ordering constant $\tilde c$
that appeared in string calculation of the tadpole in ref.\
\cite{GNL_I}; the redefinition $c = \tilde c - \ln 3$ turns out to be
convenient since then all $\gD_s$ with $s = t, d_\pm$ in
\eqref{ApprxgD} contain just a single constant term. (All functions
$\gD_s$ are expressed as a sum of untwisted propagators on the string
world sheet at zero separation. A logarithmically divergent terms has
to be subtracted in this limit, which introduces the arbitrary
constant $\tilde c$, as we have discussed in \cite{GNL_I}.)
It should also be stressed that  the approximation for
$\gD_u(\gt,\bgt)$ is exact up to order $\exp(-2\pi\, \gt_2)$; in particular
it does not contain any order $\exp(-2\pi\, \gt_2/3)$ or 
$\exp(-4\pi\, \gt_2/3)$ terms. This shows that the field theory
approximation of $\gD_u$ given in \eqref{QgDfield} is very
accurate, with the first corrections appearing at order $e^{-2 \pi 
\gt_2}$. Similarly the reason why field theory fails to
notice the exponentially decaying corrections to the twisted
propagators can easily be understood:  The field theory is expected to
be accurate only for the infra--red  region $\gt_2 \mg 1$. Apart from
the exponentially suppressed corrections to the (twisted) propagators,
we see that the propagators depend on $\gt_1$.  This dependence  
makes a distinction between the various boundary conditions in the
twisted sectors.

Apart from these stringy corrections that decay exponentially with
$\gt_2$, two  constants have appear in the expansion
\eqref{ApprxgD}: The string normal ordering constant $c$ and $-3\ln 3$. 
The latter is a consequence of the $\Intr_3$ orbifold which we are
considering throughout this work. As far as we know there are no
fundamental principles in string theory that could fix the value of
the normal ordering constant. 
However, as we observed in \cite{GNL_I}, the Fourier transform of 
the Gaussians in \eqref{SketchTadp} to configuration space 
\eqref{SketchTadpCoord} is only well--defined provided that all
functions $\gD_s$ never turn negative. This leads to a lower bound
$c_0$ for the normal ordering constant $c$. From the formulae
\eqref{ApprxgD} we find the approximate bound 
\equ{
c \geq c_0 = 6 e^{-2 \pi \,  \frac 13\, \gr} 
- 9 e^{-2 \pi \, \frac 23\, \gr} + \ldots \approx 0.75824.
\labl{NorBound}
}
(The numerical value is obtained using the exact string expressions 
evaluated at $\gr = \frac 12 \sqrt 3$.)  Moreover, from the
approximations we infer that all twisted correlators approach the
normal ordering constant $c$ for large $\gt_2$. This leads to one of
the most fundamental differences between string and field theory: 
In field theory the twisted states have constant profiles in $k$,
while in string theory they always decay exponentially (as $\exp(|k|^2)$) 
precisely because this constant is strictly positive: 
$c \geq c_0 \approx 0.75824 >0$. In addition, the function $\gD_u$
includes the constant part $c - 3 \ln 3$ which is absent in field
theory, as is apparent from comparing \eqref{ApprxgD} with
\eqref{QgDfield}.

\begin{figure}
\begin{center}
\tabu{cc}{
\includegraphics[width=7.5cm]{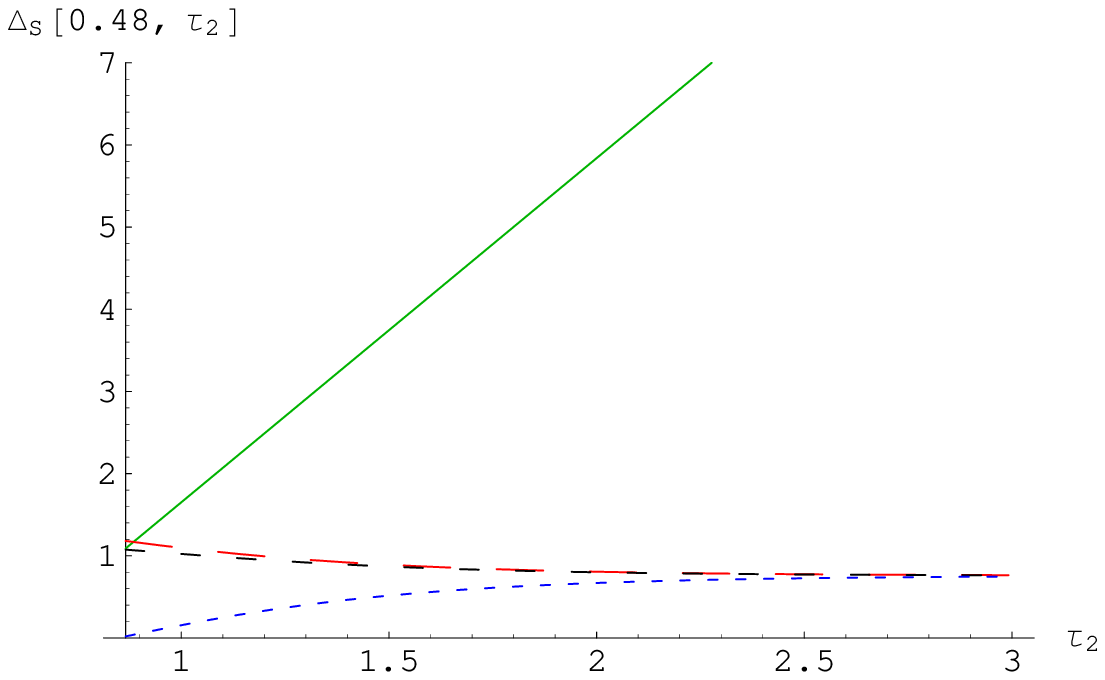} &
\includegraphics[width=7.5cm]{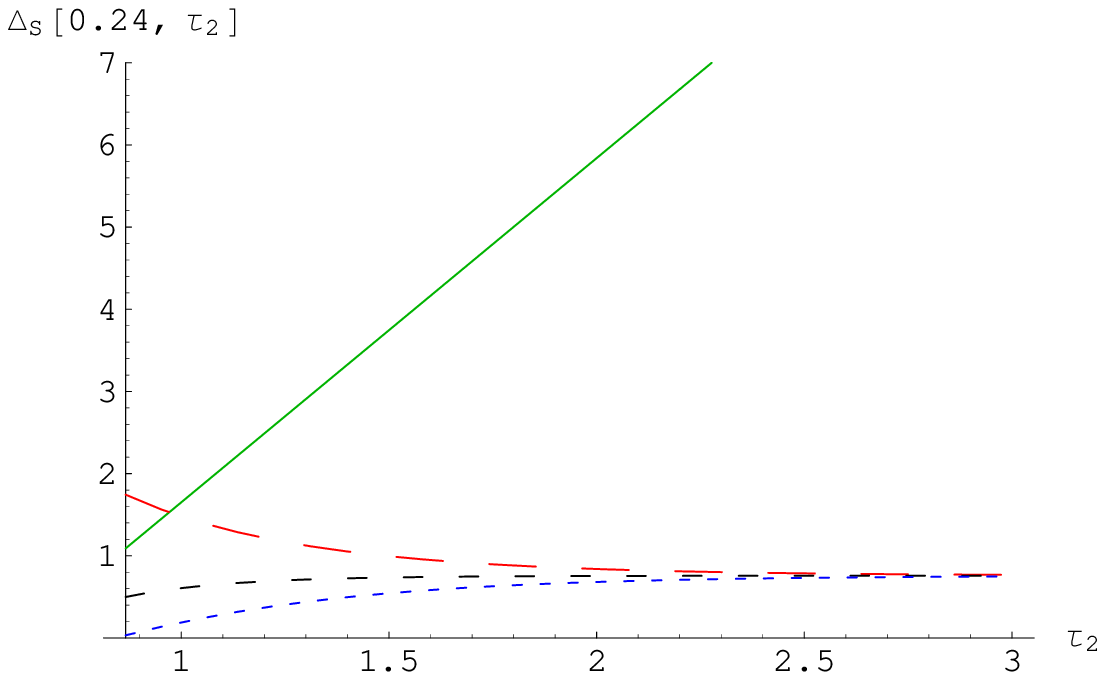} 
\\[2ex]
\includegraphics[width=7.5cm]{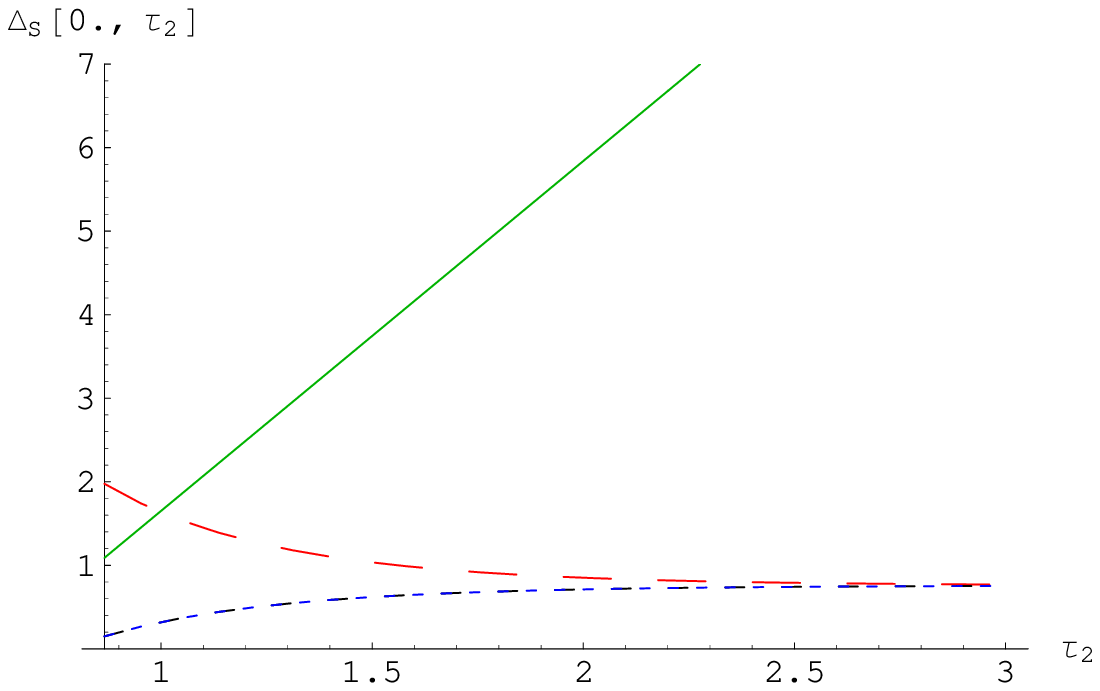} &
\includegraphics[width=7.5cm]{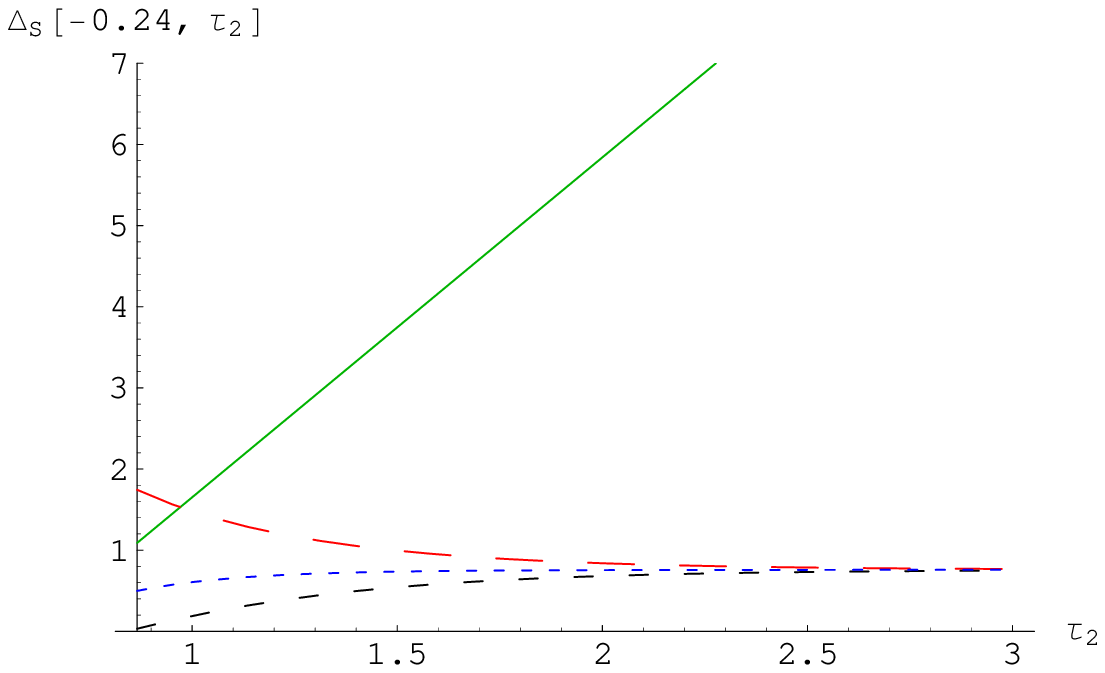} 
\\[2ex]
\includegraphics[width=7.5cm]{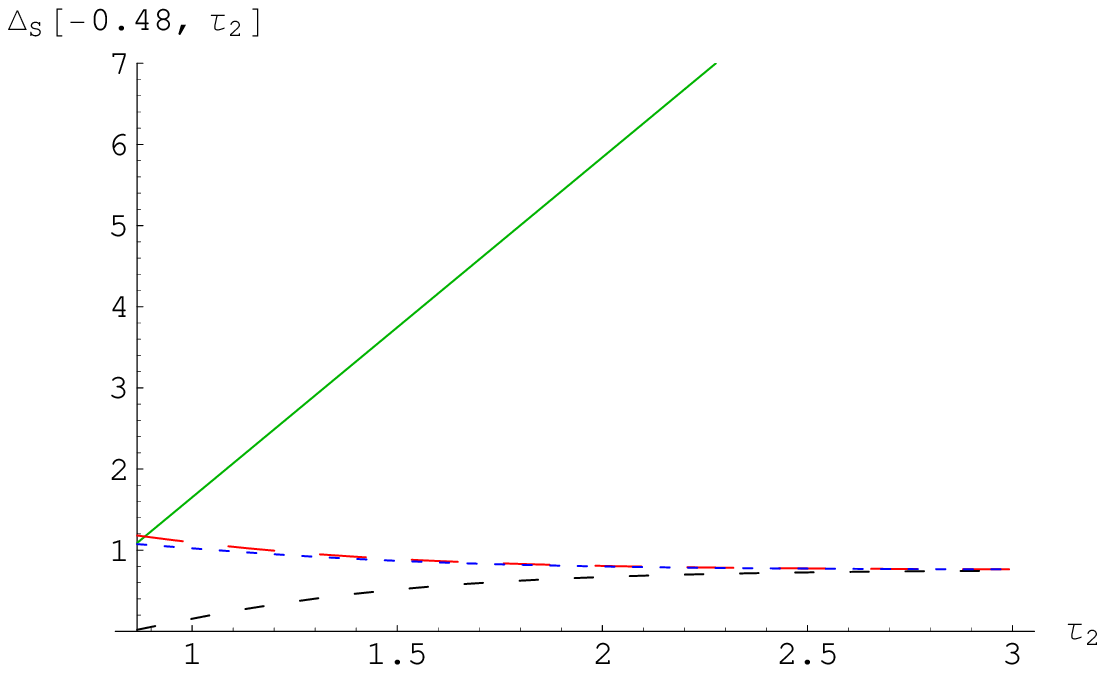} &
\raisebox{4ex}{\scalebox{0.65}{\mbox{\input{LngdgD.pstex_t}}}}
}
\end{center}
\caption{ 
The five plots show the differences of the functions 
$\gD_s(\gt_1,\gt_2)$ for the values $\gt_1 = -0.48, -0.24, 0, 0.24$ and
$0.48$ in the different sectors $s = u, t, d_+$ and $d_-$. (The values
were mainly motivated to show the changes in $\gD_s$ with $\gt_1$
clearly, and to avoid too many coincident curves.)  
The change of the function $\gD_u$ for these values of $\gt_1$ is
hardly  visible as can be seen from its approximation in equation
\eqref{ApprxgD}. The roles of $\gD_{d_+}$ and $\gD_{d_-}$ are
interchanged when we take  $\gt_1 \ra -\gt_1$. 
\labl{fg:gDplots}}
\end{figure}

We have plotted the functions $\gD_s(\gt_1, \gt_2)$ for the different
sectors $s = u, t, d_+$ and $d_-$ in figure \ref{fg:gDplots} for five 
values $\gt_1$. It is important to note that the behavior of the
untwisted states ($u$) is very different than that of the twisted
states ($t$, and  $d_\pm$) for large $\gt_2$: The former grow linearly
with $\gt_2$, while the latter all approach the constant $c_0$, as
expected from approximations \eqref{ApprxgD}. The function $\gD_t$ is
symmetric under the sign flip of $\gt_1 \ra -\gt_1$, and that 
\equ{
\gD_{d_+}(-\gt_1, \gt_2) = \gD_{d_-}(\gt_1, \gt_2).
\labl{RelationgDpm}
}
(Using the exact expressions for these functions derived in 
\cite{GNL_I}, these properties can be verified analytically.) In the plots
\ref{fg:gDplots} we have chosen the string normal ordering constant equal
to the  minimum value for which all propagators for all $\gt$ are
positive definite. If we take $c$ larger than $c_0$, 
all curves are shifted upwards by the same amount.

In direct analogy to our field theory definition \eqref{LocWidthF}, 
we introduce localization widths for the different sectors as  
\equ{
\ovr{\gD_s} = \frac{3}{2\gp}\, 
\int_\cF  \frac {\d^2 \gt}{\gt_2^2} 
\,  {\gD_s(\gt_1,\gt_2)},
\labl{LocWidthS} 
}
and the normalization factor $3/(2\gp)$ has been included to ensure
the interpretation of an average. Notice that as in the field theory
case $\ovr{\gD_u}$ is logarithmically divergent.  This is another
indication that string theory  treats the untwisted states
as genuine bulk states. Furthermore, it is obvious that larger values of 
the normal ordering constant $c$ correspond to larger localization 
widths for all fields: Larger $c$ corresponds to more delocalized 
effects  of the fixed point fields. We observe that choosing  $c$ to
take its minimum value $c_0$ gives localizations that are the closest
to the naive field theory expectation. In that case we find 
\equ{
\ovr{\gD_u} = \infty, 
\qquad 
\ovr{\gD_t} \approx 1.98627, 
\qquad 
\ovr{\gD_{d_+}} = \ovr{\gD_{d_-}} \approx 1.38752. 
\labl{SlocWidth}
} 
Since the twisted localization widths are quite close to unity, the
spread of these states is of the order of the string scale. The
hierarchy between $\ovr{\gD_u}$, $\ovr{\gD_t}$ and $\ovr{\gD_{d_\pm}}$
can be easily understood qualitatively: From either the approximate
formulae \eqref{ApprxgD} or from the plots in figure \ref{fg:gDplots}
we see that for most values of $\gt$ we have 
\(
\gD_u(\gt) > \gD_t(\gr) > \gD_{d_\pm}(\gt), 
\)
hence similar inequalities hold for their averages.

%% file: torus.pstex_t
\begin{picture}(0,0)%
\includegraphics{torus.pstex}%
\end{picture}%
\setlength{\unitlength}{3947sp}%
\begingroup\makeatletter\ifx\SetFigFont\undefined%
\gdef\SetFigFont#1#2#3#4#5{%
  \reset@font\fontsize{#1}{#2pt}%
  \fontfamily{#3}\fontseries{#4}\fontshape{#5}%
  \selectfont}%
\fi\endgroup%
\begin{picture}(2187,2067)(2026,-5758)
\put(3751,-5536){\makebox(0,0)[lb]{\smash{\SetFigFont{14}{16.8}{\rmdefault}{\mddefault}{\updefault}{\color[rgb]{0,0,0}$1$}%
}}}
\put(2026,-3886){\makebox(0,0)[lb]{\smash{\SetFigFont{14}{16.8}{\rmdefault}{\mddefault}{\updefault}{\color[rgb]{0,0,0}$i$}%
}}}
\put(2701,-4186){\makebox(0,0)[lb]{\smash{\SetFigFont{14}{16.8}{\rmdefault}{\mddefault}{\updefault}{\color[rgb]{0,0,0}$\gt$}%
}}}
\put(2326,-5686){\makebox(0,0)[lb]{\smash{\SetFigFont{14}{16.8}{\rmdefault}{\mddefault}{\updefault}{\color[rgb]{0,0,0}$\gt_1$}%
}}}
\put(2776,-4936){\makebox(0,0)[lb]{\smash{\SetFigFont{14}{16.8}{\rmdefault}{\mddefault}{\updefault}{\color[rgb]{0,0,0}$\gt_2$}%
}}}
\end{picture}

%% file: FunDom.pstex_t
\begin{picture}(0,0)%
\includegraphics{FunDom.pstex}%
\end{picture}%
\setlength{\unitlength}{3947sp}%
\begingroup\makeatletter\ifx\SetFigFont\undefined%
\gdef\SetFigFont#1#2#3#4#5{%
  \reset@font\fontsize{#1}{#2pt}%
  \fontfamily{#3}\fontseries{#4}\fontshape{#5}%
  \selectfont}%
\fi\endgroup%
\begin{picture}(2487,2667)(3589,-6661)
\put(5401,-5611){\makebox(0,0)[lb]{\smash{\SetFigFont{14}{16.8}{\rmdefault}{\mddefault}{\updefault}{\color[rgb]{0,0,0}$\gr = \frac 12 \sqrt 3$}%
}}}
\put(5176,-6661){\makebox(0,0)[lb]{\smash{\SetFigFont{14}{16.8}{\rmdefault}{\mddefault}{\updefault}{\color[rgb]{0,0,0}$\frac 12$}%
}}}
\put(4201,-6661){\makebox(0,0)[lb]{\smash{\SetFigFont{14}{16.8}{\rmdefault}{\mddefault}{\updefault}{\color[rgb]{0,0,0}-$\frac 12$}%
}}}
\put(6076,-6436){\makebox(0,0)[lb]{\smash{\SetFigFont{14}{16.8}{\rmdefault}{\mddefault}{\updefault}{\color[rgb]{0,0,0}$\gt_1$}%
}}}
\put(4726,-4186){\makebox(0,0)[lb]{\smash{\SetFigFont{14}{16.8}{\rmdefault}{\mddefault}{\updefault}{\color[rgb]{0,0,0}$\gt_2$}%
}}}
\put(4876,-5386){\makebox(0,0)[lb]{\smash{\SetFigFont{14}{16.8}{\rmdefault}{\mddefault}{\updefault}{\color[rgb]{0,0,0}$1$}%
}}}
\end{picture}

%% file: LngdgD.pstex_t
\begin{picture}(0,0)%
\includegraphics{LngdgD.pstex}%
\end{picture}%
\setlength{\unitlength}{3947sp}%
\begingroup\makeatletter\ifx\SetFigFont\undefined%
\gdef\SetFigFont#1#2#3#4#5{%
  \reset@font\fontsize{#1}{#2pt}%
  \fontfamily{#3}\fontseries{#4}\fontshape{#5}%
  \selectfont}%
\fi\endgroup%
\begin{picture}(1372,1647)(1779,-2167)
\put(3151,-736){\makebox(0,0)[lb]{\smash{\SetFigFont{17}{20.4}{\rmdefault}{\mddefault}{\updefault}{\color[rgb]{0,0,0}$\gD_u$}%
}}}
\put(3151,-2086){\makebox(0,0)[lb]{\smash{\SetFigFont{17}{20.4}{\rmdefault}{\mddefault}{\updefault}{\color[rgb]{0,0,0}$\gD_{d_-}$}%
}}}
\put(3151,-1186){\makebox(0,0)[lb]{\smash{\SetFigFont{17}{20.4}{\rmdefault}{\mddefault}{\updefault}{\color[rgb]{0,0,0}$\gD_t$}%
}}}
\put(3151,-1636){\makebox(0,0)[lb]{\smash{\SetFigFont{17}{20.4}{\rmdefault}{\mddefault}{\updefault}{\color[rgb]{0,0,0}$\gD_{d_+}$}%
}}}
\end{picture}

%% file: Fshape.tex
\section{Shape of tadpoles}
\labl{sc:shapes}

\begin{figure}
\begin{center}
\hspace{-1cm}{
\tabu{cc}{
\epsfig{file=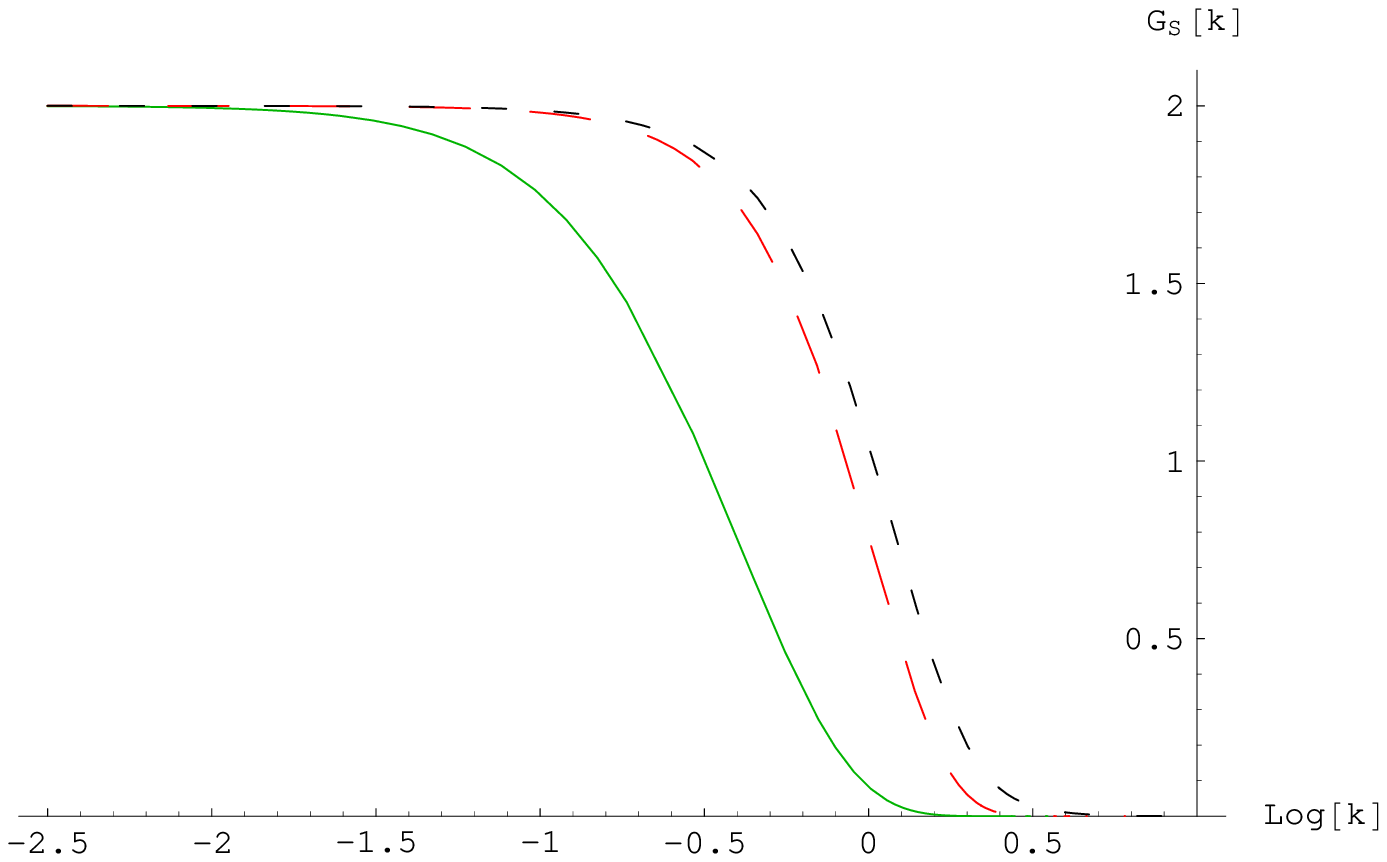,width=8cm}
&
\epsfig{file=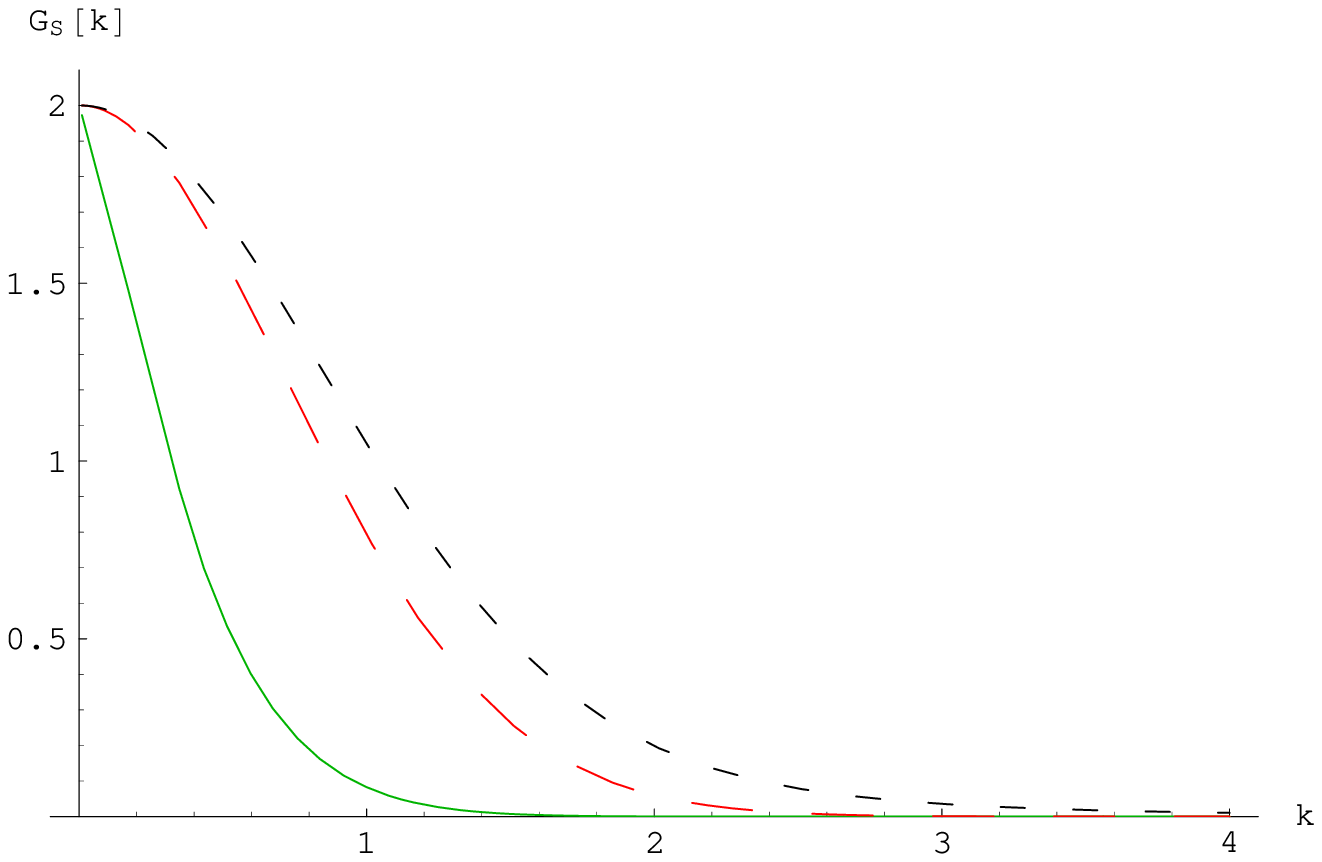,width=8cm}
\hspace{-3cm}{
\raisebox{3cm}{\scalebox{0.6}{\mbox{\input{LngdGauss.pstex_t}}}}}
}}
\end{center}
\caption{These two figures display the momentum profiles for the $u$,
$t$ and $d_\pm$--sector. The picture on the left is plotted using a
(decimal) log--scale; the picture on the right uses a linear scale.  
\labl{fg:SU9shape}}
\end{figure}

In this section we investigate the different profiles for gauge field
tadpoles that arise at the one--loop level in string theory. In
particular we wish to examine how the twisted states contribute in the
neighborhood of the orbifold singularity to give us some insight
as to how fixed point states should be treated in field theory.

The general form \eqref{SketchTadp} shows that there are two functions
that determine the profile of the tadpole, the first is the regularized 
two point function of the coordinate fields $\gD_s$, and the second is the 
charge trace $Q_s^b$ defined in \eqref{SketchTadpCoord}. In contrast
to the universal nature of these widths $\gD_s$, which only depend on
the orbifold under investigation, the charges $Q_s^b$ are very 
model dependent. For the $\Intr_3$ orbifolds we consider there are only
two different models that have anomalous $\U{1}$ and non--vanishing 
charge traces \cite{GNL_I}. We consider the $\SU{9}$ model first.  It
is the simpler of the two cases, and it will illustrate the generic 
properties of the tadpole profiles.  With the $\SU{9}$ model
understood we can contrast it with the $\E{7}$ model which exhibits
some surprising features.

\subsection*{$\boldsymbol{\SU{9}}$ model}

For the $\Intr_3$ model containing the gauge group $\SU{9}$ the
charge functions $Q_s^b$ are exactly constant and all equal within
each $\E{8}$: 
\equ{
Q_u^1 = Q_t^1 = Q_{d_+}^1 = Q_{d_-}^1 = 0, 
\qquad 
Q_u^2 = Q_t^2 = Q_{d_+}^2 = Q_{d_-}^2 = 2. 
}
We interpret  this result as implying that all contributions, excluding
the zero modes, exactly cancel each other in the loop in each
sector. This shows that for this model an effective field theory
approach to string theory, keeping only the string zero modes, takes
all contributing states into account.  We focus only on the
non--vanishing charge functions in the  second $\E{8}$.
A nice (accidental) feature of this model, which we will exploit for 
illustrative purposes, is that the charges $Q_s^2$ are all equal.
This allows us to isolate the differences between the various sectors
that arise due to the differences in the functions $\gD_s$ only.  In figure
\ref{fg:SU9shape} the momentum profiles for the untwisted ($u$) states
and twisted ($t, d_\pm$) states are plotted.  We see that the untwisted 
states are  more suppressed for higher momentum than the 
twisted states.  
The field theory approximation of the contribution from twisted states
is equal for any value of the internal momenta $k$, while the string
results always damps momenta much larger than the string scale.

\begin{figure}
\begin{center}
\hspace{0cm}{
\tabu{ccc}{
\epsfig{file=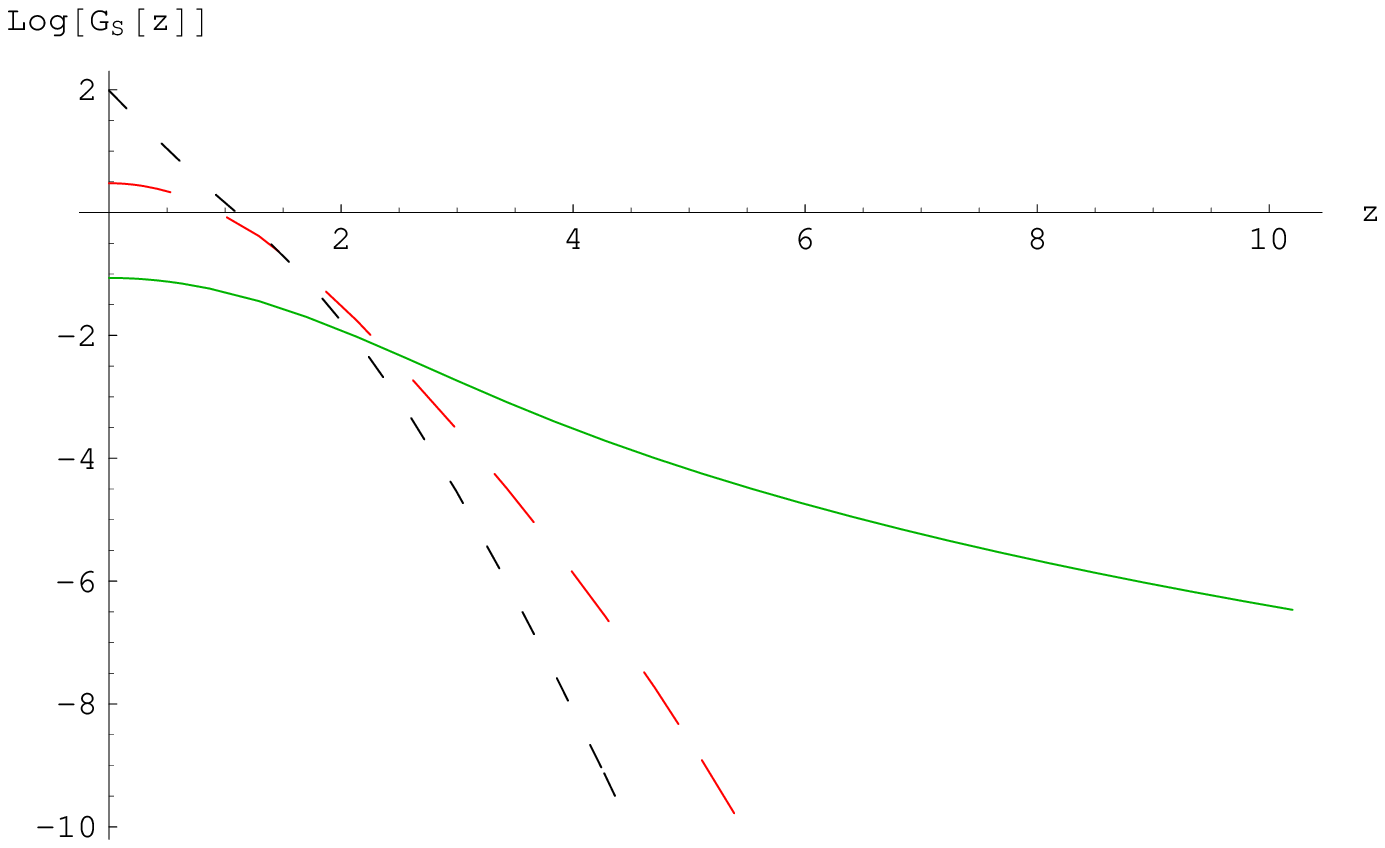,width=10cm}
& & 
\hspace{0cm}{
\raisebox{4.6cm}{\scalebox{0.6}{\mbox{\input{LngdGauss.pstex_t}}}}}
}}
\end{center}
\caption{The spatial extension of the tadpole contributions in the
sectors $u, t$ and $d_\pm$ are displayed on a logarithmic scale.  
The curve for the $u$--sector fall off much slower than the others for
large values of $|z|$, as only it corresponds to bulk contributions.
\labl{fg:SU9shapeCoor}}
\end{figure}

In the coordinate space representation we find more pronounced 
differences between the drop off in the sectors. In figure
\ref{fg:SU9shapeCoor} we have plotted the profiles as a function of
the radial variable $|z|$ in the six internal dimensions. Close to the
orbifold singularity the twisted states in the $t$ and
$d_\pm$--sectors dominate the tadpole. At a distance of about 
$z = 2.5$ string length all three sectors contribute with comparable
magnitude. After that the curves corresponding to the $t$ and
$d_\pm$--sectors fall of much faster than the one for the
$u$--sector. These differences are consistent with our understanding
that the $u$--sector corresponds to bulk states, while the $t$ and
$d_\pm$--sectors constitute the fixed point states.

The difference between the plots can be understood because equation
\eqref{SketchTadpCoord} contains factors $1/\gD_s$ both in the overall
normalization and in the exponential. This implies that the smallest
values of $\gD_s$ will contribute the most to the tadpoles in the
coordinate space representation. However, a smaller value of the
function $\gD_s$ will give a much stronger suppression  from the
exponential for sizable values of $|z|$. Since $\gD_u$ is the largest
of the functions $\gD_s$, as can be verified by inspecting the plots
\ref{fg:gDplots}, the profile $\cG_u(z)$ is least suppressed  for
large values of $|z|$. The situation close to the orbifold singularity
($|z| \approx 0$), is the opposite, because then the exponential factor 
approaches unity. The values of $\cG_s$ there are determined by the
normalizations $1/\gD_s^3$, hence close to the singularity the sector
$d_\pm$  dominates.

We can also investigate how the profiles of the tadpoles in the
coordinate representation are affected by the value of the normal
ordering constant.  We have plotted these profiles for  a range of $c$
in figure  \ref{fg:SU9shapeCoorNor}.  
The overall effects are that close to the orbifold singularity, all the
profiles drop significantly for large values of $c$. When $|z|$
becomes of the order of two to four string lengths, the curves are
falling off less fast for larger values of the normal ordering
constant. For very large $|z|$ the contribution of the untwisted
sector become identical irrespective of what the normal ordering
constant is.

These features can be understood by noting that all the
curves in figure  \ref{fg:SU9shapeCoorNor} correspond to normalized
Gaussians integrated over the fundamental domain: When integrated over
$\Cplx^3/\Intr_3$ all curves give the same value (which is equal to
$2$ in the $\SU{9}$ model).  From their
definitions \eqref{SketchTadpCoord} it follows that for larger values
of the normal ordering constant, the tadpole are smaller because
the factor $1/\gD_s^3$ gives a bigger suppression near $z=0$, where
the exponent can be neglected. But since these profiles are
normalized, they need to become larger when $|z|$ is increased. Hence,
as we would have expected from the localization widths
\eqref{SlocWidth}, the profiles of the twisted states drop off slower
for the larger value of $c$.  A similar argument applies to the
untwisted states: Since their localization width is infinite, the 
integral over the orbifold receives significant contributions for 
any value of $|z|$. Therefore to respect the normalization condition
all untwisted curves need to tend to each other for large $|z|$. This
analysis provides a different argument that taking $c = c_0$ leads to 
physics that most closely resembles the field theory treatment of
the twisted states, as they are the most localized to the
orbifold fixed points.

\begin{figure}
\begin{center}
\tabu{c}{
\tabu{ccc}{
$u$--sector & $t$--sector & $d_\pm$--sector
\\[0ex]
\epsfig{file=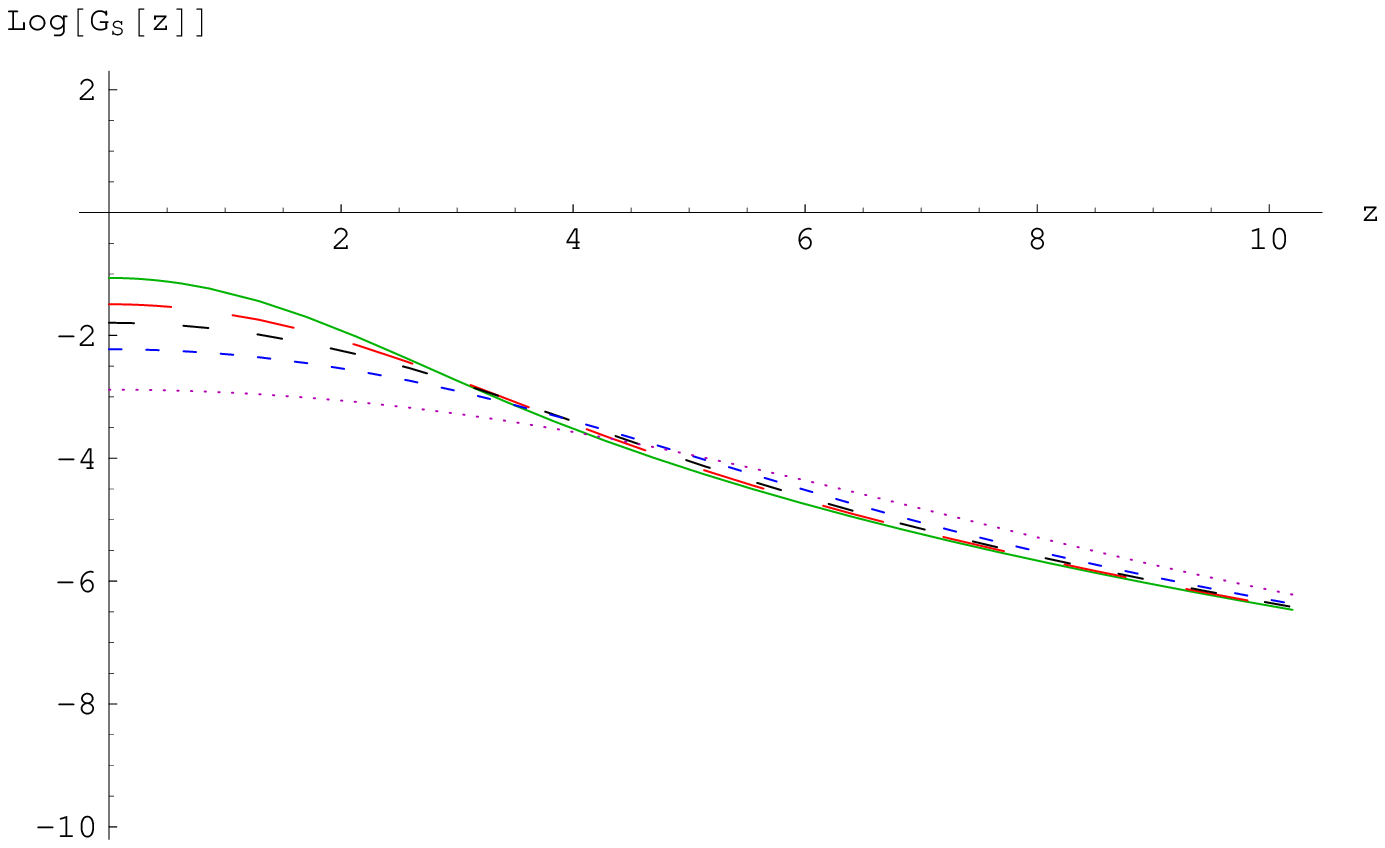,width=5cm}
&
\epsfig{file=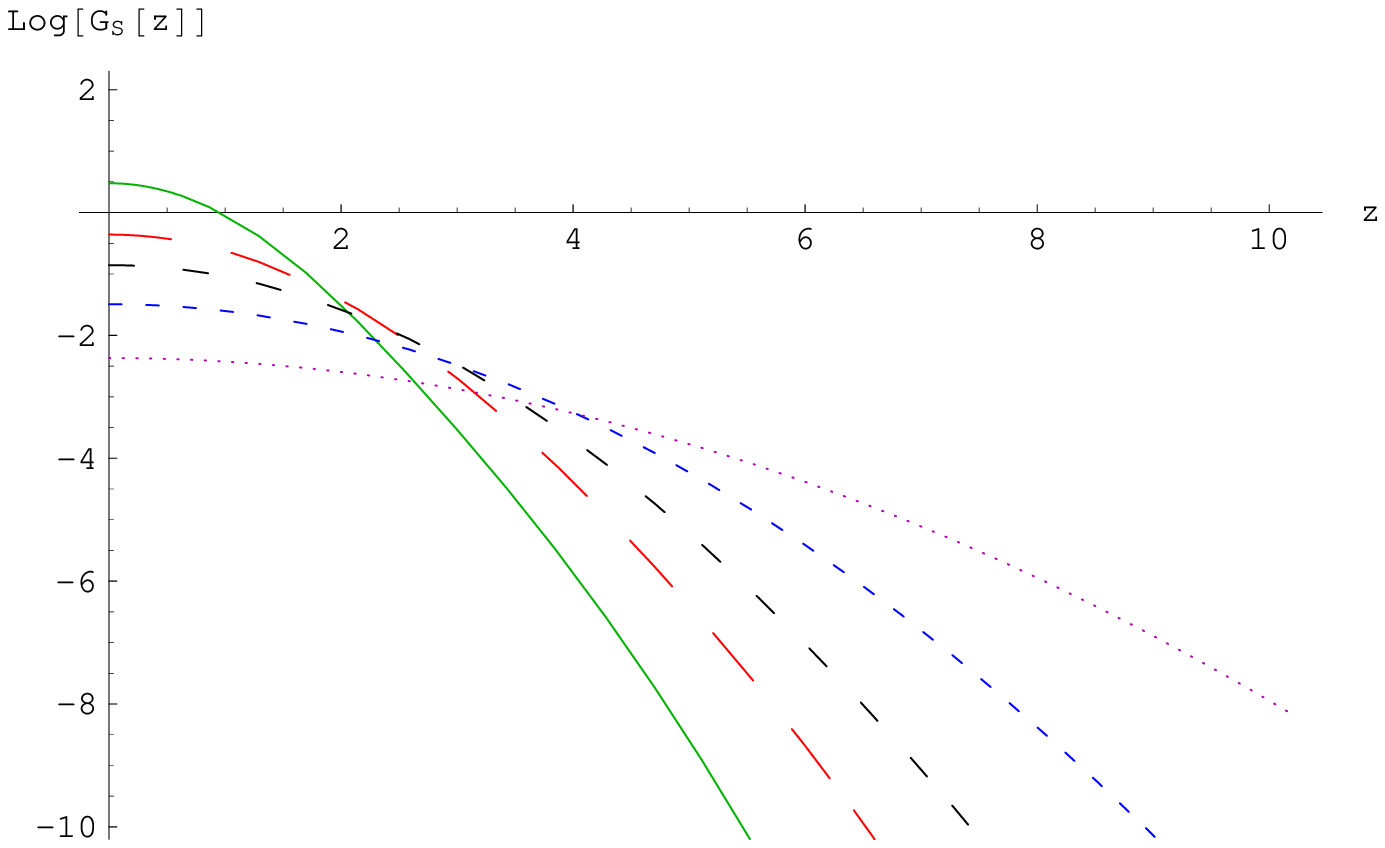,width=5cm}
&
\epsfig{file=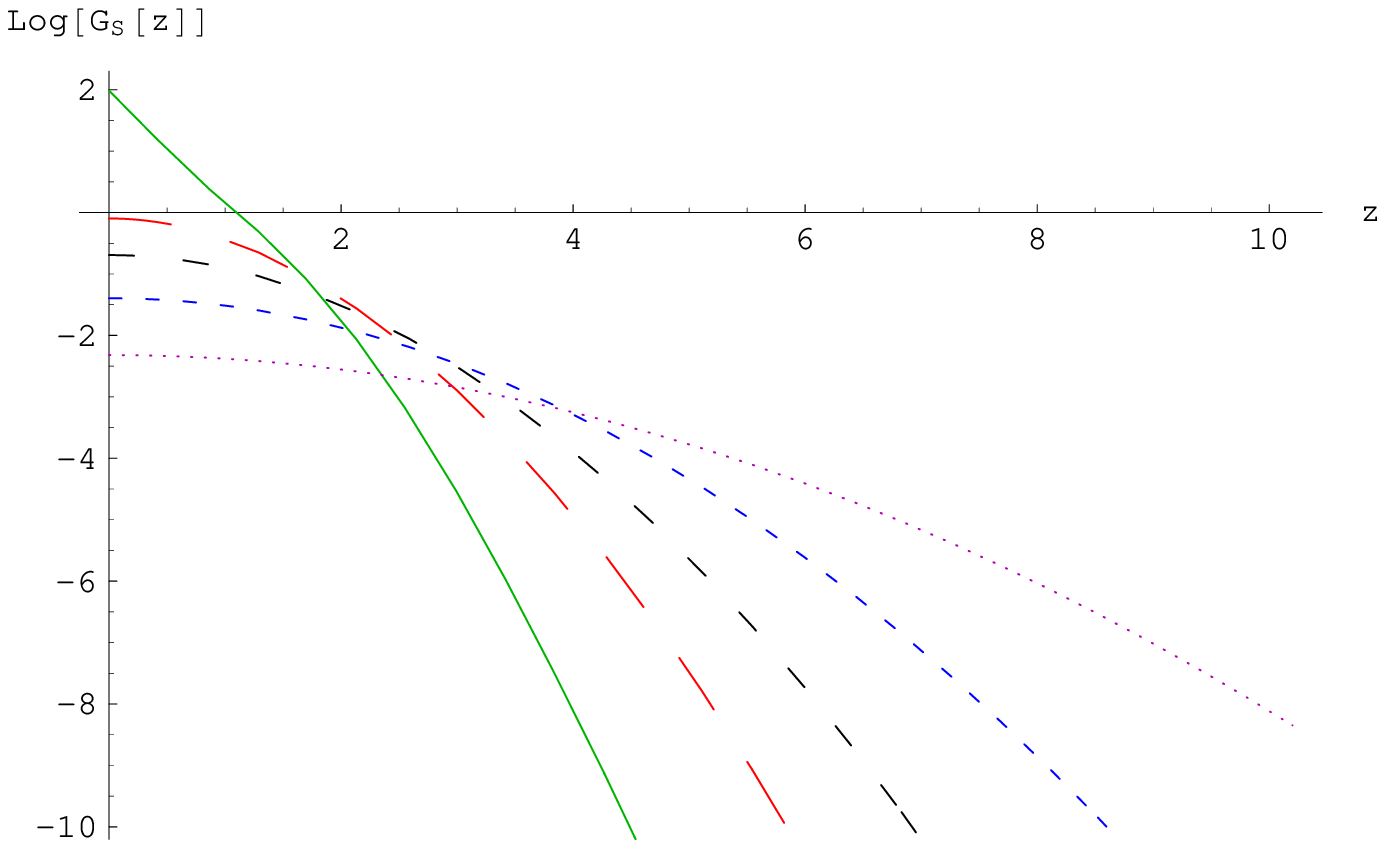,width=5cm}
}
\\[2ex]
\hspace{0cm}{
\raisebox{0cm}{\scalebox{0.6}{\mbox{\input{LngdVar.pstex_t}}}}}
}
\end{center}
\caption{In this plot we compare the shapes tadpoles that arise when
we take the normal ordering constant $c$ equal to 
$c_0, 2c_0,3c_0, 5c_0$ or $10c_0$, with $c_0$ the minimum value of the
normal ordering constant \eqref{NorBound}.  
We see that the larger the value of $c$ the more the profiles of the
different sectors are suppressed for small values of $|z|$, while at
the same time the twisted state contributions become more
delocalized. 
\labl{fg:SU9shapeCoorNor}}
\end{figure}

\subsection*{$\boldsymbol{\E{7}}$ model}

Next, we compare the situation of the $\SU{9}$ model with the other
anomalous model: The $\E{7}$ has two $\U{1}$ factors in its zero mode
gauge group, see table \ref{tab:z3models}.  The expressions for
$Q_s^b$ for the $\E{7}$ model are more complicated than those of the
$\SU{9}$ model and we give their expansions up to order 
$q = \exp(2 \pi i\, \gt)= \exp(2 \pi i \gt_1) \exp( - 2 \pi \gt_2)$ 
\equ{
\arry{l}{
Q_u^1~~ \approx ~~ 3 + \ldots, 
\\[1ex] 
Q_t^1 ~~\approx ~~ 
\frac{1}{9}\,  q^{-\frac{1}{3}} + \frac{5}{3} + 6\,q^{\frac{1}{3}} - 
   \frac{76}{9} \,q^{\frac{2}{3}} + \ldots, 
\\[1ex]
Q_{d_+}^1 \approx 
~~ \frac{1}{9}\, e^{i \frac{4\pi}{3}} \,q^{-\frac{1}{3}} + \frac{5}{3} 
+ 6\,e^{i\frac{2\pi}{3}}\,q^{\frac{1}{3}} 
- \frac{76}{9}\, e^{i \frac{4 \pi}{3}} \, q^{\frac{2}{3}} + \ldots, 
\\[1ex]
Q_{d_-}^1 \approx 
~~  \frac{1}{9}\,e^{i\frac{2\pi}{3}}\, q^{-\frac{1}{3}} +  \frac{5}{3} 
+ 6\,e^{i \frac{4 \pi}{3}}\, q^{\frac{1}{3}} 
- \frac{76}{9}\, e^{i \frac{2\pi}{3}} \, q^{\frac{2}{3}} + \ldots, 
}
\labl{Qexp1}
}
and 
\equ{
\arry{l}{
Q_u^2 ~~ \approx ~~  2 + \ldots,  
\\[1ex]
Q_t^2 ~~ \approx ~~ 
  \frac{2}{9}\,q^{-\frac{1}{3}} - \frac{2}{3} + 12\,q^{\frac{1}{3}} - 
   \frac{152}{9} \,q^{\frac{2}{3}} + \ldots, 
\\[1ex] 
Q_{d_+}^2 \approx 
 ~~ \frac{2}{9} \,e^{i \frac{4\pi}{3}} \,q^{-\frac{1}{3}} -  \frac{2}{3} 
+ 12\, e^{i \frac{2\pi}{3}}\, q^{\frac{1}{3}} 
- \frac{152}{9}\, e^{i\frac{4 \pi}{3}}\, q^{\frac{2}{3}} + \ldots, 
\\[1ex]
Q_{d_-}^2 \approx 
~~ \frac{2}{9} \, e^{i\frac{2\pi}{3}} \, q^{-\frac{1}{3}} - \frac{2}{3}  
+ 12\, e^{i \frac{4 \pi}{3}}\, q^{\frac{1}{3}} 
- \frac{152}{9}\, e^{i \frac{2\pi}{3}}\, q^{\frac{2}{3}} + \ldots. 
}
\labl{Qexp2}
} 
The negative power $q^{- 1/3}$ indicate that tachyonic states
contribute to the charge functions $Q^b_s$ for the twisted sectors 
$s = t, d_+$ and $d_-$. The positive powers result from massive string
excitations. In fact a tower of massive string states contribute to
these local shapes. This constitutes the most profound difference
between the $\SU{9}$ and $\E{7}$ models:

In the $\E{7}$ model the tachyonic and massive string excitations give
non--vanishing contributions, so the effective field theory
calculation of the local profile of tadpole, which takes only the
zero modes into account, is therefore questionable. The approximation
that ignores the massive states  can be justified because of the
suppression factor of at least $\exp ( - 2 \pi \gt_2/3)$. In contrast 
the tachyonic contribution are enhanced by the factor 
$\exp(2 \pi \gt_2/3)$ so they could well give the leading effect. We will
see below that they actually dominate the tadpole close to the
orbifold fixed point, showing that effective field theory with zero
modes ignores the most significant contributions near the orbifold
singularity.  For this comparison we do not need to compute
the zero mode contributions again, since they are simply given by the
profiles discussed in the previous subsection rescaled by the
appropriate charges, which can be read  from the constant
terms in \eqref{Qexp1} and \eqref{Qexp2}.

To investigate to what extent the tachyonic contributions to the
tadpoles constitute the leading effect, we begin by defining the
tachyonic momentum profile of the tadpole by 
\equ{
T(k) = 
\int_\cF  \frac {\d^2 \gt}{\gt_2^2} 
\, e^{- 2 \pi i\, \frac {\gt_1}3} 
\, e^{2 \pi \, \frac {\gt_2}3 }
\left\{ 
e^{- \gD_t(\gt) \, \bar k k }
+ 
e^{i \frac {2 \pi}3} \, 
e^{- \gD_{d_-}(\gt) \, \bar k k }
+ 
e^{i \frac {4 \pi}3} \, 
e^{- \gD_{d_+}(\gt) \, \bar k k }
\right\}. 
\labl{TachTadp}
}
This is obtained by identifying the part of \eqref{SketchTadp} that is 
proportional to $q^{-1/3} =\exp ( - 2 \pi i \gt /3 )$ for the $\E{7}$ model.
We must  treat the tachyonic contributions of all twisted
($t, d_+$ and $d_-$) sectors together a finite result: The integral over
$\gt_2$ is divergent for each separately because  the respective 
integrands are unbounded for large $\gt_2$. As the functions
$\gD_s(\gt)$ with $s = t, d_\pm$ all approach the normal ordering
constant $c$ for large $\gt_2$ (see the approximations \eqref{ApprxgD}
or figures \ref{fg:gDplots}) the contributions cancel in this limit,
by  virtue of the $\Intr_3$ projector identity  
\equ{
\frac 13 \Bigl( 1 + e^{2 \pi i/3} + e^{4 \pi i/3} \Bigr) = 0. 
}
This relation also implies that the tachyonic contributions
vanish in the limit in which only the zero modes contribute, i.e.\ 
$k \ra 0$.  As the zero mode spectrum does not contain tachyons, the
vanishing of $T(k)$ in this limit is the correct behavior.  
This suggests that it is more accurate to say that the tachyons cancel 
each other at the zero mode level, rather than that they are entirely 
absent. One might worry that because tachyonic states run around in
the loop, one can describe the decay of a gauge field in two tachyons
by cutting the diagram using the optical theorem.\footnote{We are
indebted to T.\ Gherghetta for raising this point.} This is not the
case because, as observed in section \ref{sc:stringtadp}, the tadpole
is real, hence its imaginary component associates this decay is zero.

\begin{figure}
\begin{center}
\epsfig{file=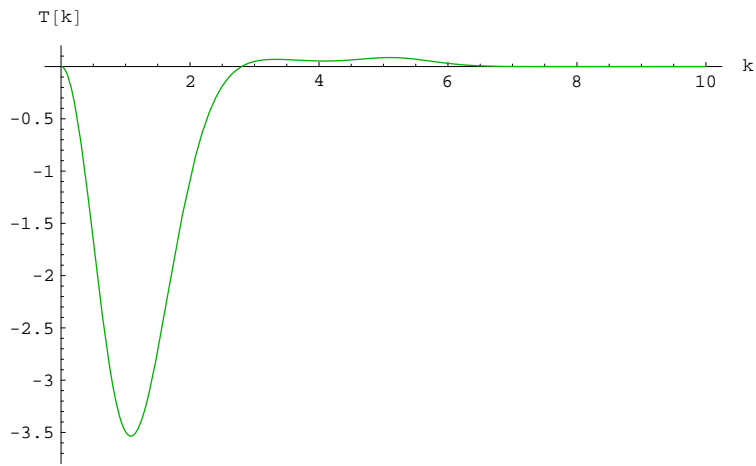,width=10cm}
\end{center}
\caption{
This plot shows the tachyonic contributions to the gauge field tadpole
as a function of the absolute value of the internal momentum $k$. 
\labl{fg:MomTach}}
\end{figure}

This limiting behavior of the tachyonic contributions to the tadpole
are readily verified in plot of figure \ref{fg:MomTach}.  To
understand the other major features of the shape of this tadpole, it
is useful to rewrite equation \eqref{TachTadp} in a manifestly real
form
\equ{
T(k) = 
\int_\cF  \frac {\d^2 \gt}{\gt_2^2} 
\, e^{2 \pi \, \frac {\gt_2}3 }
\left\{ 
\cos \bigl( \frac {2\pi}3 \gt_1 \bigr) 
\left( 
e^{- \gD_t(\gt) \, \bar k k }
- 
E_+(k| \gt) \right) 
- \sqrt 3\, \sin \bigl( \frac {2\pi}3 \gt_1 \bigr) E_-(k | \gt) 
\right\}. 
\labl{TachTadpReal}
}
We have introduced the even/odd functions 
\equ{
E_\pm(k|\gt) = \frac 12 
\Bigl( 
e^{- \gD_{d_+}(\gt) \, \bar k k }
\pm 
e^{- \gD_{d_-}(\gt) \, \bar k k }
\Bigr), 
\qquad 
E_\pm(k| -\gt_1, \gt_2) = \pm E_\pm(k | \gt_1, \gt_2). 
} 
Using \eqref{RelationgDpm} it is straightforward to verify the parity 
under $\gt_1 \rightarrow -\gt_1$, and that $E_+(0|\gt) = 1$ and 
$E_-(0|\gt) = 0$. Close to $k = 0$ the term with $E_-$ gives a
negligible contributions compared to the first two terms. 
We know that $\gD_t > \gD_{d_\pm}$ and so $E_+(k|\gt)$ is bigger
than the first term in \eqref{TachTadpReal} for non--zero $k$. 
Consequently, the integral first becomes more and more
negative. However, for larger $k$ $E_-(k|\gt)$ becomes comparable to
$E_+(k|\gt)$. As the integrals give the largest contributions where
the propagators are the smallest, i.e.\  $\gt_1 \approx \pm 1/2$, the
$E_-(k|\gt)$ contributions dominates because of the factor 
$\sqrt 3\, \sin(2 \pi \gt_1/3) \approx \pm 3/2$. 
This means that the integrand turns positive, and for large values of 
$k$ the integral takes positive values. As  $k$ increases the
exponential damping suppressed all contributions, and $T(k)$ tends to
zero.

It is also instructive to study the tadpole due to the tachyons in the
coordinate space representation. This tadpole can also be cast in a
manifestly real form analogous to the expression
\eqref{TachTadpReal}: 
\equ{
\cT(z) = 
\int_\cF  \frac {\d^2 \gt}{\gt_2^2} 
\, e^{2 \pi \, \frac {\gt_2}3 }
\left\{ 
\cos \bigl( \frac {2\pi}3 \gt_1 \bigr) 
\Bigl( 
\frac{e^{- |z|^2/  \gD_t(\gt) } }{ (\gp \gD_t(\gt) )^3 } 
- 
\cE_+(k| \gt) \Bigr) 
- \sqrt 3\, \sin \bigl( \frac {2\pi}3 \gt_1 \bigr) \cE_-(k | \gt) 
\right\}. 
\labl{CoorTachTadpReal}
}
Like the exponential factor, the even and odd functions 
\equ{
\cE_\pm(z|\gt) = \frac 12 
\Bigl( 
\frac{ e^{- |z|^2/ \gD_{d_+}(\gt)} }{(\gp \gD_{d_+}(\gt))^3 } 
\pm 
\frac{ e^{- |z|^2/ \gD_{d_-}(\gt)} }{(\gp \gD_{d_-}(\gt))^3} 
\Bigr), 
\qquad 
\cE_\pm(z| -\gt_1, \gt_2) = \pm \cE_\pm(z| \gt_1, \gt_2),
} 
contain the six dimensional Gaussian normalization factors 
$1/(\gp \gD_{d_\pm})^3$. Therefore, the integral over the orbifold of
each of these exponentials is normalized to unity. It follows that the
integrated tadpole due to tachyons vanish. This reflects the
cancellation of the tachyonic contributions within the zero mode theory.

\begin{figure}
\begin{center}
\epsfig{file=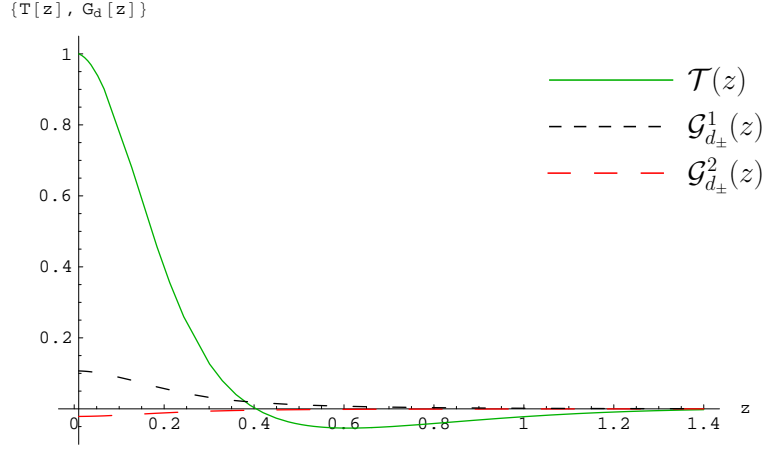,width=10cm}
\put(-80,100){\scalebox{0.65}{\mbox{\input{LngdTachZero.pstex_t}}}}
\end{center}
\caption{
The coordinate space representation of the tachyonic contributions to
the gauge field tadpole is displayed by the solid line. This
contribution  dominates the largest zero mode contribution
coming from the $d_\pm$ sectors. 
\labl{fg:CoorTach}}
\end{figure}

The full coordinate profile of the tachyonic contributions to the
tadpole is depicted in figure \ref{fg:CoorTach} by the solid  green line. 
The main features can be explained by a similar analysis as performed
above, but now taking the properties of the Gaussian normalization
factor into account. For the behavior near $z=0$ this is already
crucial, because contrary to the case above, $\cE_-(z|\gt)$ does not
vanish but is dominated by $1/(\gp \gD_{d_\pm}(\gt))^3$ with 
$\gt \approx (\mp1 + i \sqrt 3)/2$. Since there 
$\sqrt 3 \, \sin(2 \pi \gt_1/3) \approx \mp 3/2$, the term with
$\cE_-(z|\gt)$ is the largest in \eqref{CoorTachTadpReal} and
positive. Since $1/\gD_t(\gt)$ is always smaller than $1/\gD_{d_\pm}$ the
first exponential factor in the expression for the coordinate
representation of the tadpole becomes the most important for large
values of $|z|$. For even larger $|z|$ the exponential
suppression forces the shape to approach zero.

In figure \ref{fg:CoorTach} the tachyonic contributions to the tadpole
is plotted together with the leading contributions from the zero
modes. Inspecting figure  \ref{fg:SU9shapeCoor} we infer that the
twisted sectors $d_{\pm}$ give the largest contributions to the zero
mode states. (The other sectors contributions are at least one
order of magnitude less, and can therefore safely be ignored in the
present analysis.) The normalization of the zero modes in
the $d_\pm$ sectors compared to the tachyonic sector is $15$ and $-3$
for the first and second $E_8$, respectively, see equations
\eqref{Qexp1} and \eqref{Qexp2}. These relative
normalizations have been taken into account in comparison figure 
\ref{fg:CoorTach}. We conclude that the tachyonic states
totally dominate  the profile of the tadpoles for the $\U{1}$'s in
both $\E{8}$'s near the singularity.

%% file: LngdGauss.pstex_t
\begin{picture}(0,0)%
\includegraphics{LngdGauss.pstex}%
\end{picture}%
\setlength{\unitlength}{3947sp}%
\begingroup\makeatletter\ifx\SetFigFont\undefined%
\gdef\SetFigFont#1#2#3#4#5{%
  \reset@font\fontsize{#1}{#2pt}%
  \fontfamily{#3}\fontseries{#4}\fontshape{#5}%
  \selectfont}%
\fi\endgroup%
\begin{picture}(1372,1251)(1779,-1717)
\put(3151,-1186){\makebox(0,0)[lb]{\smash{\SetFigFont{17}{20.4}{\rmdefault}{\mddefault}{\updefault}{\color[rgb]{0,0,0}$G_t$}%
}}}
\put(3151,-1636){\makebox(0,0)[lb]{\smash{\SetFigFont{17}{20.4}{\rmdefault}{\mddefault}{\updefault}{\color[rgb]{0,0,0}$G_{d_\pm}$}%
}}}
\put(3151,-736){\makebox(0,0)[lb]{\smash{\SetFigFont{17}{20.4}{\rmdefault}{\mddefault}{\updefault}{\color[rgb]{0,0,0}$G_u$}%
}}}
\end{picture}

%% file: LngdVar.pstex_t
\begin{picture}(0,0)%
\includegraphics{LngdVar.pstex}%
\end{picture}%
\setlength{\unitlength}{3947sp}%
\begingroup\makeatletter\ifx\SetFigFont\undefined%
\gdef\SetFigFont#1#2#3#4#5{%
  \reset@font\fontsize{#1}{#2pt}%
  \fontfamily{#3}\fontseries{#4}\fontshape{#5}%
  \selectfont}%
\fi\endgroup%
\begin{picture}(10725,280)(-449,-512)
\put(10276,-436){\makebox(0,0)[lb]{\smash{\SetFigFont{17}{20.4}{\rmdefault}{\mddefault}{\updefault}{\color[rgb]{0,0,0}$10 c_0.$}%
}}}
\put(8101,-436){\makebox(0,0)[lb]{\smash{\SetFigFont{17}{20.4}{\rmdefault}{\mddefault}{\updefault}{\color[rgb]{0,0,0}$5 c_0,$}%
}}}
\put(6001,-436){\makebox(0,0)[lb]{\smash{\SetFigFont{17}{20.4}{\rmdefault}{\mddefault}{\updefault}{\color[rgb]{0,0,0}$3c_0,$}%
}}}
\put(3676,-436){\makebox(0,0)[lb]{\smash{\SetFigFont{17}{20.4}{\rmdefault}{\mddefault}{\updefault}{\color[rgb]{0,0,0}$2 c_0,$}%
}}}
\put(1576,-436){\makebox(0,0)[lb]{\smash{\SetFigFont{17}{20.4}{\rmdefault}{\mddefault}{\updefault}{\color[rgb]{0,0,0}$c_0,$}%
}}}
\put(-449,-436){\makebox(0,0)[lb]{\smash{\SetFigFont{17}{20.4}{\rmdefault}{\mddefault}{\updefault}{\color[rgb]{0,0,0}$c =$}%
}}}
\end{picture}

%% file: LngdTachZero.pstex_t
\begin{picture}(0,0)%
\includegraphics{LngdTachZero.pstex}%
\end{picture}%
\setlength{\unitlength}{3947sp}%
\begingroup\makeatletter\ifx\SetFigFont\undefined%
\gdef\SetFigFont#1#2#3#4#5{%
  \reset@font\fontsize{#1}{#2pt}%
  \fontfamily{#3}\fontseries{#4}\fontshape{#5}%
  \selectfont}%
\fi\endgroup%
\begin{picture}(1387,1251)(1764,-1717)
\put(3151,-1186){\makebox(0,0)[lb]{\smash{\SetFigFont{17}{20.4}{\rmdefault}{\mddefault}{\updefault}{\color[rgb]{0,0,0}$\cG_{d_\pm}^1(z)$}%
}}}
\put(3151,-1636){\makebox(0,0)[lb]{\smash{\SetFigFont{17}{20.4}{\rmdefault}{\mddefault}{\updefault}{\color[rgb]{0,0,0}$\cG_{d_\pm}^2(z)$}%
}}}
\put(3151,-736){\makebox(0,0)[lb]{\smash{\SetFigFont{17}{20.4}{\rmdefault}{\mddefault}{\updefault}{\color[rgb]{0,0,0}$\cT(z)$}%
}}}
\end{picture}

%% file: Fconcl.tex
\section{Discussion}
\labl{sc:concl}

We devote the final section of this paper to a discussion of our
string theory results for local gauge field tadpoles. We speculate as to
how some of the stringy effects we encountered could be included in
field theory.

In this work we have studied the structure of gauge field tadpoles in
heterotic string on the (non--compact) orbifold $\Cplx^3/\Intr_3$, as
recently computed in \cite{GNL_I}. The tadpoles are given as an
integral over the fundamental domain of a sum of  charge functions
times Gaussian distributions. The charge functions encode the 
states that contribute to the tadpole.   By contrast the Gaussian
distributions are universal in the sense that they affect the
profiles of all participating states. The widths of these Gaussian
distributions are determined by the expectation value of the square of
the internal coordinate fields on the string world sheet. Since the
shape of the fundamental domain dictates that $\gt_2 > \sqrt 3/2$,
the expansions in fractional powers of $\exp(-2 \pi \gt_2)$ are rapidly
convergent.

By keeping only the zeroth order in this expansion of these Gaussian
widths, we recovered the behavior of untwisted and twisted states in
field theory up to the addition of some constants. For the twisted
states these constants are particularly important, since they
lead to a smooth cut--off for large internal momenta. This is in contrast
to the naive field theory treatment of fixed point states which
assumes that the contribution of any  internal momentum is
equally. When looking at the tadpole in a coordinate space
representation, the effect of these constants is that the twisted
states are not localized like delta functions at the fixed point, but
rather are spread out around the orbifold singularity with widths
comparable to a few string lengths. The string normal ordering
constant has set a  regularization scale for the orbifold singularity.

Because of the clear physical importance of these constants, we
suggest that such constants should always be included in a field
theoretical description of fixed point states. The transition from the string
theory to the field theory treatment can be easily achieved by
replacing the integration over the fundamental domain by an
integration over Schwinger proper time. This methodology 
gives a well defined cut--off procedure because the string normal 
ordering constant acts as a regulator, that manifestly preserves
Lorentz invariance.

A natural question therefore is whether these constants can be
determined from first principals. In string theory these constants have 
the interpretation of normal ordering constants, which arise because
the subtraction of singular terms on the string world sheet has an
ambiguity.  As we showed in \cite{GNL_I} all tadpole widths depend on
a single normal ordering constant. We have observed that this constant
has to bounded from below. If this bound is violated, the target space
description of the string theory breaks down because the Gaussian
widths may become negative for some values of the modular
parameters. We were not able to establish an upper bound on this
constant, however, we have argued that the localization of the twisted
states is the strongest if the lower bound is saturated.  This most
closely mimics the naive field theory description of the states at the
fixed points.

These arguments show that we have some constraints on these 
constants that set the width of the twisted states over the internal 
dimensions from string theory. We doubt that direct field theoretical
principles will allow us say much about them, because it is precisely 
the functional dependence on $\exp(-2 \pi \gt_2)$ of the widths that led
to the lower bound in the string theory analysis.  We find that the $\gt_1$ 
dependence of the widths $\gD_s$ have the effect of increasing the lower 
bound for $c$.

As described above, the Gaussian distributions are only one
aspect of the structure of the 
one--loop gauge field tadpoles. They also contain `charges' which are
functions of the modular parameter as well: Massless states give 
constant contributions, while massive string excitations give rise to
some positive power of  $\exp(-2\pi \gt_2)$. However, unlike the
widths of the Gaussian distributions, these are not the only 
corrections: In the model in  which  $\E{8} \times \E{8}'$ is broken to  
$\E{7} \times \U{1} \times \SO{14}' \times \U{1}'$, we found that
there are also tachyons contributing with a growing exponential
$\exp(2 \pi \gt_2/3)$. This does not constitute a contradiction with the
statement that heterotic string theories on orbifold do not contain
tachyons in their spectrum: The tachyons contributing to the tadpole
are not physical particles as they  only exist as off--shell quantum
corrections in the loop. 
Moreover, we showed that when the
tadpole is integrated over the orbifold, the tachyonic contributions
cancel each other. As figure \ref{fg:CoorTach} showed they only give
contributions in the vicinity of the orbifold singularity, but there
they completely dominate all other massless and massive string
contributions!

This remarkable result that tachyons in some string models can give
sizable contributions to local operators, like gauge field tadpoles,
is somewhat provocative. Normally, by considering the zero modes of the
string, one assumes that one captures the major part of the
perturbative quantum corrections.  Here we see that for local
physics the tachyons sometimes play the most  important role. 
In string theory these tachyonic contributions can be
identified as straightforwardly as the massless or massive string
excitations. On the other hand, in field theory it is not all obvious
when these local tachyonic contributions should be included, as it is
not simply the presence of an anomalous $U(1)$ which requires that
they should be considered. Therefore, further research on local  
tachyonic corrections is necessary to gain a more complete
understanding of their relevance.

\section*{Acknowledgements} 

We would like to thank T.\ Gherghetta for useful discussions. 
The work of S.G.N.\ was partially supported by 
DOE grant DE--FG02--94ER--40823.  The work of M.L.\ was 
partially supported by NSERC Canada.  S.G.N.\ would like to thank the
University of British Columbia for their hospitality during which this
project got started. M.L.\ wishes to thank the University 
of Minnesota for their hospitality during part of the preparation of this 
work.